\documentclass[sigconf]{acmart}
\renewcommand\footnotetextcopyrightpermission[1]{}
\usepackage{caption, subcaption}
\usepackage{booktabs}
\usepackage{multirow}
\usepackage{pdfpages}
\usepackage{balance}
\usepackage{etoolbox}

\AtBeginDocument{
  \providecommand\BibTeX{{
    \normalfont B\kern-0.5em{\scshape i\kern-0.25em b}\kern-0.8em\TeX}}}

\copyrightyear{2022}
\acmYear{2022}
\setcopyright{acmcopyright}\acmConference[MM '22]{Proceedings of the 30th ACM International Conference on Multimedia}{October 10--14, 2022}{Lisboa, Portugal}
\acmBooktitle{Proceedings of the 30th ACM International Conference on Multimedia (MM '22), October 10--14, 2022, Lisboa, Portugal}
\acmPrice{15.00}
\acmDOI{10.1145/3503161.3547880}
\acmISBN{978-1-4503-9203-7/22/10}

\begin{document}

\title{High-Fidelity Variable-Rate Image Compression via Invertible Activation Transformation}

\author{Shilv Cai}

\affiliation{
	\institution{Huazhong University of Science and Technology}
	\streetaddress{1037 Luoyu Road, Hongshan District}
	\city{}
	\state{}
	\country{}
	\postcode{430074}
}
\email{caishilv@hust.edu.cn}

\author{Zhijun Zhang}
\affiliation{
	\institution{Huazhong University of Science and Technology}
	\streetaddress{1037 Luoyu Road, Hongshan District}
	\city{}
	\state{}
	\country{}
	\postcode{430074}
}
\email{zhangzhijun@hust.edu.cn}

\author{Liqun Chen}
%\authornote{Corresponding author.}
\affiliation{
	\institution{Huazhong University of Science and Technology}
	\streetaddress{1037 Luoyu Road, Hongshan District}
	\city{}
	\state{}
	\country{}
	\postcode{430074}
}
\email{chenliqun@hust.edu.cn}

\author{Luxin Yan}
\affiliation{
	\institution{Huazhong University of Science and Technology}
	\streetaddress{1037 Luoyu Road, Hongshan District}
	\city{}
	\state{}
	\country{}
	\postcode{430074}
}
\email{yanluxin@hust.edu.cn}

\author{Sheng Zhong}
\affiliation{
	\institution{Huazhong University of Science and Technology}
	\streetaddress{1037 Luoyu Road, Hongshan District}
	\city{}
	\state{}
	\country{}
	\postcode{430074}
}
\email{zhongsheng@hust.edu.cn}

\author{Xu Zou}
\affiliation{
	\institution{Huazhong University of Science and Technology}
	\streetaddress{1037 Luoyu Road, Hongshan District}
	\city{}
	\state{}
	\country{}
	\postcode{430074}
}
\email{zoux@hust.edu.cn}
\renewcommand{\shortauthors}{Shilv Cai et al.}

%%
%% By default, the full list of authors will be used in the page
%% headers. Often, this list is too long, and will overlap
%% other information printed in the page headers. This command allows
%% the author to define a more concise list
%% of authors' names for this purpose.
%\renewcommand{\shortauthors}{Trovato and Tobin, et al.}

%% The abstract is a short summary of the work to be presented in the
%% article.
\begin{abstract}
Learning-based methods have effectively promoted the community of image compression.
Meanwhile, variational autoencoder~(VAE) based variable-rate approaches have recently gained much attention to avoid the usage of a set of different networks for various compression rates.
Despite the remarkable performance that has been achieved, these approaches would be readily corrupted once multiple compression/decompression operations are executed, resulting in the fact that image quality would be tremendously dropped and strong artifacts would appear~(see Figure~\ref{fig:teaser}).
Thus, we try to tackle the issue of high-fidelity fine variable-rate image compression and propose the Invertible Activation Transformation~(IAT) module.
We implement the IAT in a mathematical invertible manner on a single rate Invertible Neural Network~(INN) based model and the quality level~(QLevel) would be fed into the IAT to generate scaling and bias tensors. 
IAT and QLevel together give the image compression model the ability of fine variable-rate control while better maintaining the image fidelity.
Extensive experiments demonstrate that the single rate image compression model equipped with our IAT module has the ability to achieve variable-rate control without any compromise. And our IAT-embedded model obtains comparable rate-distortion performance with recent learning-based image compression methods. 
Furthermore, our method outperforms the state-of-the-art variable-rate image compression method by a large margin, especially after multiple re-encodings.
\end{abstract}

%%
%% The code below is generated by the tool at http://dl.acm.org/ccs.cfm.
%% Please copy and paste the code instead of the example below.
%%
%\begin{CCSXML}
%<ccs2012>
%<concept>
%<concept_id>10010147.10010178</concept_id>
%<concept_desc>Computing methodologies~Artificial intelligence</concept_desc>
%<concept_significance>500</concept_significance>
%</concept>
%</ccs2012>
%\end{CCSXML}
%\ccsdesc[500]{Computing methodologies~Artificial intelligence}


\maketitle

\begin{figure*}
	\centering
	\includegraphics[width=0.92\linewidth,  height=6.3cm]{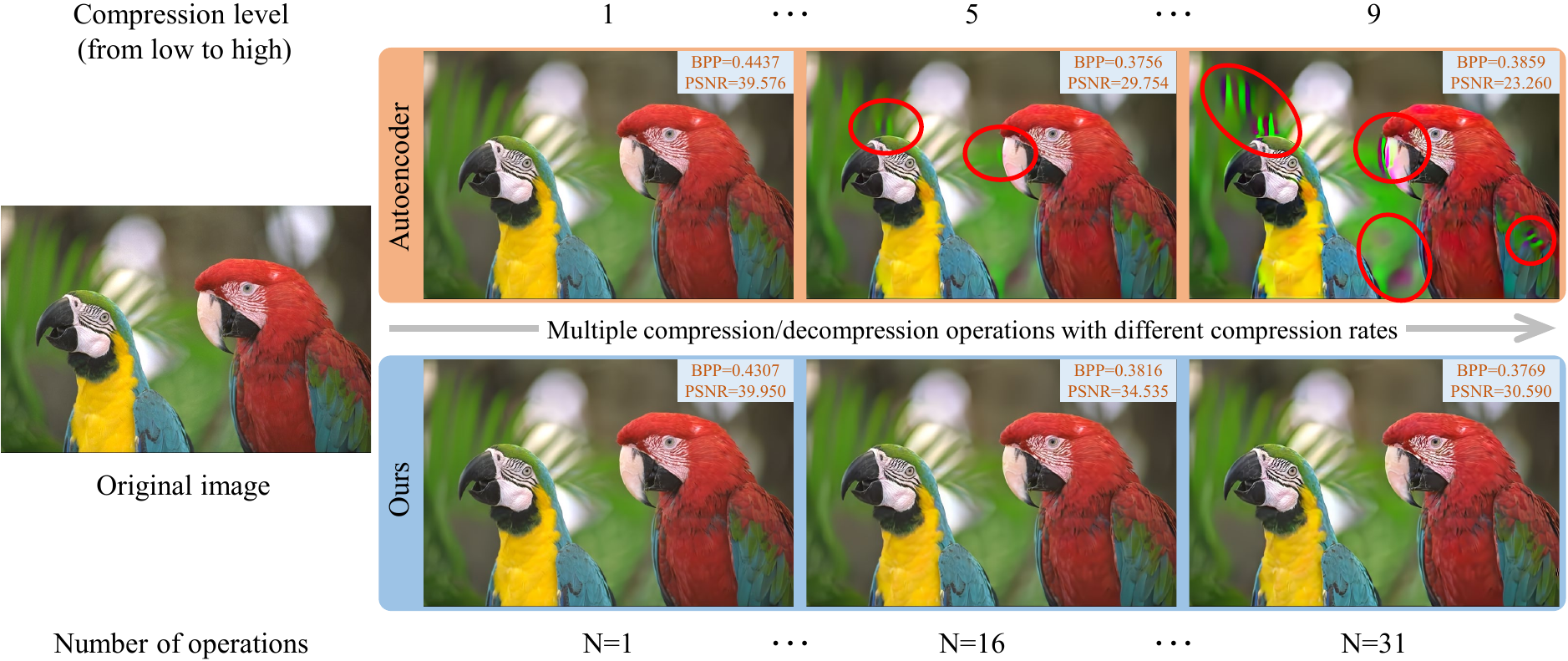}
	\caption{
		Reconstructed images of variable-rate compression methods after different numbers of compression/decompression operations. It broadly occurs in multimedia transmissions among social platforms.
		Severe artifacts and color shifts would appear~(see regions in \textcolor{red}{red circles}) in the state-of-the-art VAE-based approach~\cite{song2021variable} once multiple re-encodings are executed, in contrast to fewer artifacts and higher fidelity results achieved by our proposed approach. High-fidelity maintenance with fine variable-rate control is the main advantage and novelty of our work.
	}
	\label{fig:teaser}
\end{figure*}

\section{Introduction}
Lossy image compression is one crucial technology due to the increasing volume of visual data in such a multimedia explosion era. This task aims at lowering data redundancy while maintaining visual fidelity and supporting efficient data storage and transmissions.
To this end, many classical image compression standards~(\textit{e.g.}, JPEG~\cite{wallace1992jpeg},  JPEG2000~\cite{rabbani2002jpeg2000}, Webp~\cite{webp}, BPG~\cite{bpg}, and Versatile Video Coding (VVC)~\cite{vvc}) have been proposed and widely used in practical applications.
Recently, learning-based image compression methods have started to show superiority in terms of common metrics, \textit{e.g.} PSNR and MS-SSIM. These methods~\cite{balle2016end, theis2017lossy, balle2018variational} make use of the powerful nonlinear transformation capability of DNNs, and perform end-to-end learning by a large number of high-quality images with a rate-distortion cost.
However, despite the exciting progress, the learning-based image compression still remains challenging once variable-rate compression adaptation is needed. Most of them require training multiple single-rate models for different rates, resulting in a high cost of model storage and training.
\par
To remedy the issue, a large number of VAE-based variable-rate image compression methods~\cite{choi2019variable,yang2020variable,chen2020variable,sun2021interpolation,cui2021asymmetric,song2021variable} have been proposed.
The researchers first try to achieve discrete rate adaptation using one single model. Choi et al.~\cite{choi2019variable} introduced conditional convolution and achieved variable rate through two-stage training. Yang et al.~\cite{yang2020variable} proposed the modulated autoencoder and achieved discrete adjustable compression rates by different Lagrange multipliers. Chen et al.~\cite{chen2020variable} inserted a set of scaling factors directly before the quantizer to achieve the discrete adjustable compression rates.
However, the performance of these methods would be dropped when conducting finer variable-rate control.
Thus, the topic of fine rate adaptation has attracted more attention recently.
Sun et al.~\cite{sun2021interpolation} obtained continuously adjustable compression rate by linear interpolation.
Cui et al.~\cite{cui2021asymmetric} achieved continuous compression rate control by exponential interpolation.
Song et al.~\cite{song2021variable} conditioned on quality map and achieved the variable rate, which requires semantic segmentation labels for training.
Though these methods have the ability of fine variable-rate compression control, they need additional gain modules or semantic labels to maintain the performance.
\par
Besides, it would be particularly interesting for a variable-rate compression model if the fidelity of images could be maintained while being transmitted multiple times between numerous entities under various compression rates, especially in the current multimedia society~(\textit{e.g.} one person may download a compressed image from Instagram and then send it to his friend via WhatsApp under another re-encoding).
However, state-of-the-art VAE-based variable-rate approaches~(\textit{e.g.} Song et al.~\cite{song2021variable}) would be readily corrupted once multiple compression/decompression operations are executed, resulting in the fact that image quality would be tremendously dropped. Strong artifacts and color shifts would appear, as shown in Figure~\ref{fig:teaser}.
The main reason is that the autoencoder transforms the image to a low-dimensional latent space and irreversibly discards information before quantization, imposing an implicit limitation on the reconstruction quality. 
To alleviate information loss, Invertible Neural Networks~\cite{helminger2020lossy, xie2021enhanced} have gained much attention to effectively preserve fidelity. 
It is worth noting that VAE-based variable-rate methods cannot be directly fused into the INN-based framework since implementations of their variable-rate control do not satisfy the bijective mapping property.
Nevertheless, there is no research on INN-based variable-rate methods to the best of our knowledge. Inspired by this, we construct a variable-rate image compression model which can maintain the fidelity, especially after multiple re-encodings, by exploring the invertibility.
\par
To sum up, we propose an Invertible Activation Transformation (IAT) module based on the INN framework. This module exhibits a mathematical invertible property to avoid discarding any information in the latent space to maintain high fidelity. Notice that it is the initial work to extend the mathematical invertibility to the variable-rate image compression. Moreover, the proposed image compression method attempts to achieve finer control of multiple variable rates, by presenting a compatible tensor-based Lagrange multiplier to train the whole model.
The contributions of our proposed method are 3-folded:
\begin{itemize}
	\item We propose an effective yet neat framework, equipped with the INN-based Invertible Activation Transformation (IAT) module, to achieve the high fidelity of reconstructed images, especially after multiple variable-rate image compression/decompression operations, in a mathematical invertible manner. This issue is rarely investigated so far.
	\item The proposed model tuned rate-distortion loss and achieved fine variable-rate control through the quality level.
%	\item The proposed model tunes the rate-distortion loss and introduces a novel tensor-based Lagrange multiplier to finer control various rates, conditioned on quality levels.
	\item Extensive experiments demonstrate the superiority of our proposed methods in rate-distortion performance, fidelity maintenance, and fine rate adaptation over three datasets, including Kodak~\cite{kodak}, CLIC~\cite{CLIC2020}, and DIV2K~\cite{Agustsson_2017_CVPR_Workshops}.
\end{itemize}

\begin{figure*}[t]
	\centering
	\includegraphics[width=\linewidth, height=5.6cm]{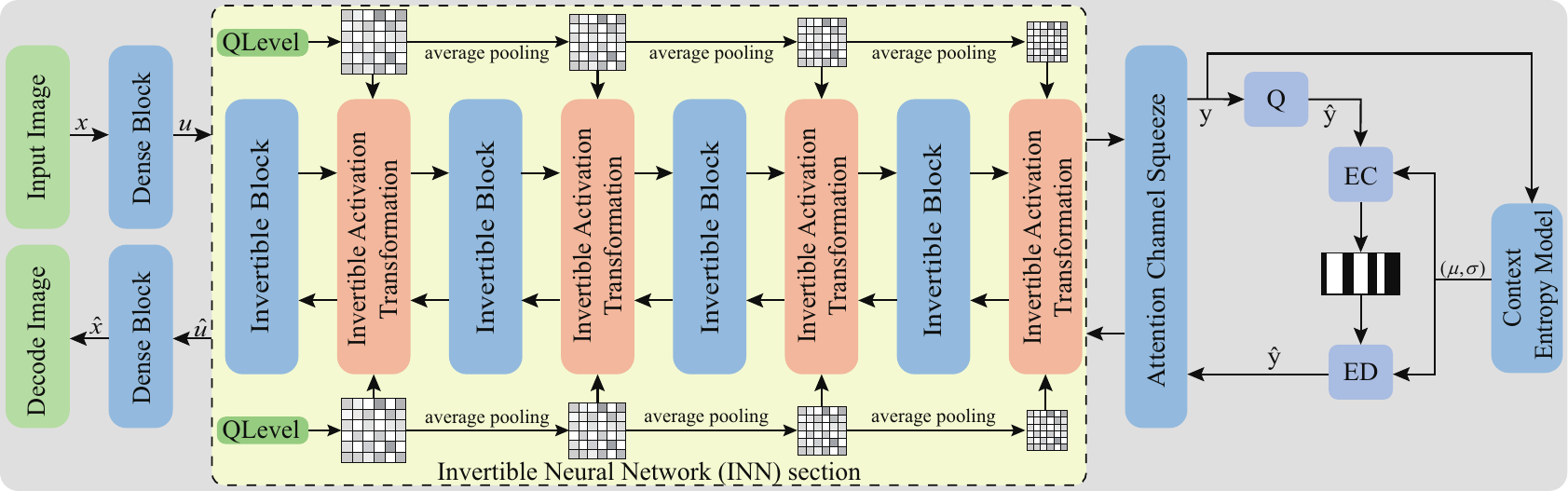}
	\caption{Network architecture equipped with the proposed Invertible Activation Transformation~(IAT) module. We insert IAT into the Invertible Neural Network section and utilize it to generate element-wise activation parameters of features from the input quality level~(QLevel). IAT and QLevel together give the image compression model the ability of fine variable-rate control while maintaining the image fidelity especially when multiple compression/decompression operations are executed.
		EC/ED means entropy encoding/decoding respectively. Q is the quantizer.
	}
	\Description{Framework_architecture}
	\label{Framework_architecture}
\end{figure*}

\section{Related Work}
In recent years, the application of neural networks in image compression has attracted widespread attention.
The variational autoencoder (VAE)~\cite{balle2016end, theis2017lossy, balle2018variational, lee2018context, mentzer2018conditional, minnen2018joint, guo20203, hu2020coarse, hu2021learning, cheng2020learned, minnen2020channel, zhou2019end, chen2021end, ma2021afec},
Invertible Neural Network (INN)~\cite{ma2020end, xie2021enhanced, helminger2020lossy, ho2021anfic}
and Generative Adversarial Networks (GAN)~\cite{rippel2017real, agustsson2019generative, iwai2021fidelity, wu2020gan, mentzer2020high} based methods have achieved surprising results.

\subsection{Learned Single Rate Image Compression}
The VAE-based framework is used as a nonlinear transformation coding model, which is the main approach in the learned image compression method. The works~\cite{balle2016end, theis2017lossy, balle2018variational} were the first to use CNN for end-to-end image compression and inspired many learning-based image compression methods.
The work~\cite{balle2018variational} introduced a hyperprior entropy model to capture the zero-mean Gaussian distribution of the latent representations. The works~\cite{minnen2018joint, lee2018context} used the Gaussian model with the non-zero mean to improve the ability to model latent representations.
Later works~\cite{lee2018context, mentzer2018conditional, minnen2018joint} further removed redundancy in potential features using the context model.
Further, the 3D-context entropy model~\cite{guo20203}, multi-scale hyperprior entropy model~\cite{hu2021learning}, and discretized Gaussian mixture model~\cite{cheng2020learned} were used to further improve the entropy model.
In addition, channel-wise module~\cite{minnen2020channel}, attention module~\cite{cheng2020learned, zhou2019end}, and non-local attention module~\cite{zhang2019residual, chen2021end} were used to extract better latent representations.
Recently transformer was used to capture long-range dependencies in probability distribution estimation effectively and efficiently~\cite{qian2022entroformer}.
\par
Most learning-based image compression methods need to train different network models for various compression rates, which not only increases the storage computational resources but also is not compatible with practical applications. Therefore, using one single model to achieve variable rate adaptation was widely studied.

\subsection{Learned Variable Rate Image Compression}
Initially, LSTM networks~\cite{toderici2015variable, toderici2017full, Johnston_2018_CVPR} control different compression rates by the different number of iterations. The more iterations, the clearer the reconstructed image would be. However, the LSTM-based approach cannot outperform JPEG2000~\cite{rabbani2002jpeg2000} in rate-distortion performance and would not obtain continuous compression rates. In addition, the iterative procedure is very time-consuming and thus not suitable for practical applications.
Then Choi et al.~\cite{choi2019variable} introduced conditional convolution in the autoencoder framework to achieve variable-rate adaptation with a single model through two-stage training.
\par
However, while variable rate is achieved, the rate-distortion performance degrades and there is a dilemma in choosing the appropriate Lagrange multiplier and quantization step size for forward inference.
Yang et al.~\cite{yang2020variable} proposed a modulated autoencoder that achieves discrete adjustable compression rate with a single model by different Lagrange multipliers.
Thesis et al.~\cite{theis2017lossy} first trained the model with high bits per pixel(bpp) and then fixed the network model parameters to train the scaling parameters for different compression rates. However, the network model suffered from incongruity with the scaling parameters, especially in low bpp cases.
Chen et al.~\cite{chen2020variable} inserted a set of scaling factors directly before the quantizer to achieve the discrete variable compression rate.
\par
Recently, research has been conducted on continuous compression rate adjustable~\cite{sun2021interpolation, cui2021asymmetric, song2021variable}. The work~\cite{cui2021asymmetric} introduced a series of vector pairs for coarse compression rate control, and then achieve continuous compression rate control by exponential interpolation.
Sun et al.~\cite{sun2021interpolation} extended the work~\cite{choi2019variable}, which obtained a continuously adjustable compression rate by linear interpolation.
Song et al.~\cite{song2021variable} conditioned the quality map by spatial feature transform (SFT)~\cite{wang2018recovering} to control different compression rates.
\par
VAE-based variable-rate approaches have been extensively researched. However, those methods suffer from severe information distortion after multiple operations of compression/decompression for the same image.
The distortion becomes more explicit as the number of operations increases.

\subsection{Invertible Neural Networks}
Invertible neural networks (INNs) are generative models that transform complex distributions into simple ones, allowing for accurate and efficient probability density estimation.
INNs have a bijective mapping of input and output, which is ideal for image compression.

NICE~\cite{dinh2014nice} introduced a flexible architecture that can learn highly nonlinear bijective transformations to represent data with simple distributions.
Based on NICE~\cite{dinh2014nice}, RealNVP~\cite{dinh2016density} further extended the idea of hierarchical and combinatorial transformations, which used affine coupling and a multi-scale framework.
Kingma et al.~\cite{kingma2018glow} proposed a generative flow model based on a $ 1 \times 1 $ invertible convolutional network with a significant improvement in log-likelihood on a standard benchmark dataset, having the advantages of exact controllability of log-likelihood, the tractability of exact inference of latent representations, and parallelizability of training and synthesis.
Ardizzone et al.~\cite{ardizzone2018analyzing} demonstrated that the validity of INNs is suitable not only for synthetic data but also for two practical applications in medicine and astrophysics.
SRFlow~\cite{lugmayr2020srflow} has designed a conditional normalizing flow architecture to solve the ill-posed problem in the super-resolution task.
Xiao et al.~\cite{xiao2020invertible} proposed an invertible rescaling network (IRN), which constructs a bijective transform to effectively implement the reconstruction of low-resolution images into high-resolution images.

INN greatly alleviates the information loss problem for better image compression, as in~\cite{ma2020end, xie2021enhanced, helminger2020lossy, ho2021anfic}. But no one has specifically studied variable-rate image compression with a single model based on the INN framework.

\section{Methodology}
\subsection{Framework}
Our image compression approach is depicted in Figure~\ref{Framework_architecture}.
%We adopt the network in Xie et al.~\cite{xie2021enhanced} as our basic architecture and introduce Invertible Activation Transformation modules to construct a single model that has the ability of fine variable-rate control while maintaining the image fidelity, especially after multiple compression/decompression operations.
% Figure \ref{Framework_architecture} illustrates the entire proposed architecture.
The proposed method implements fine variable-rate modulation in an invertible neural network framework, which involves the invertible activation transformation (IAT) module to control different compression rates through different quality levels. 
We present the detailed procedure of the model in the following:
Firstly, the source image $x \in \mathbb{R}^{3\times H\times W}$ is enhanced by the dense block module~\cite{huang2017densely} to generate a nonlinear representation of $u \in \mathbb{R}^{3\times H\times W}$, where $H$ and $W$ denote the height and width of the input image respectively. Then the forward pass of the Invertible Neural Network section, which is equipped with the proposed IAT module, transforms $u$ to a latent representation, conditioned on the quality level $L \in \mathbb{R}^{H\times W}$ to control the compression rate. This latent representation would be further fed into the Attention Channel Squeeze module to reduce the number of channels and obtain the potential representation $y$. This procedure could be formulated by a parametric analysis transform function, \textit{i.e.},
\begin{equation}
	y=g_{a}(x, L),
\end{equation}
the discrete latent features $\hat{y}$ are obtained by quantification of $y$, \textit{i.e.}, $\hat{y} = Q(y)$. We use the quantizer $Q(\cdot)$ in Ballé et al.~\cite{balle2018variational} to model the quantized latent representation $\hat{y}$ approximately by adding the uniform noise $U(-0.5, 0.5)$ to the latent representation $y$ during training and rounding the latent representation $y$ during testing. 
The context entropy model generates parameters $\mu$ and $\sigma$ of the Gaussian entropy model that approximates the distribution of quantified latent representation $\hat{y}$ to support the entropy encoding. We use range asymmetric numeral system~\cite{duda2013asymmetric} to losslessly compress latent representation $\hat{y}$ and $\hat{z}$ into bitstreams.

The inverse calculation takes the quantified latent representation $\hat{y}$ and the quality level $L$ as the input, and reconstructs the decompressed images by a parametric synthesis transform, which is formulated as follows:
\begin{equation}
	\hat{x}=g_{s}(\hat{y}, L).
\end{equation}

\subsection{Invertible Activation Transformation}
We proposed the invertible activation transformation (IAT) module to enhance the invertible neural network, which efficiently generates the desired compressed representation conditional on the quality level $L$.
The proposed IAT module can achieve variable-rate adaption on a single model while maintaining the image fidelity, especially after multiple compression/decompression operations, in a mathematical invertible manner.
\par
\begin{figure}[t]
	\centering
	\includegraphics[width=0.85\linewidth, height=4.1cm]{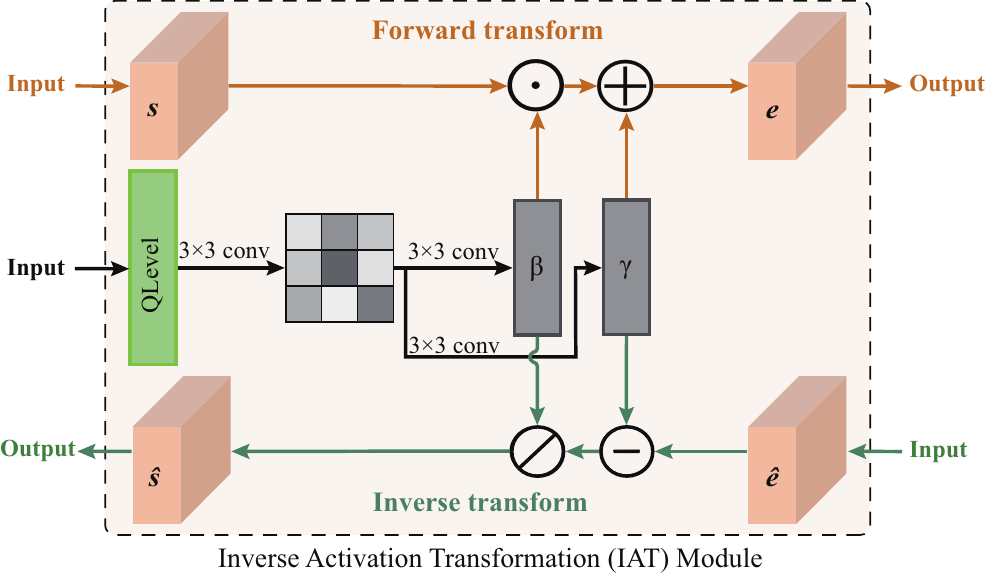}
	\caption{Illustration of the IAT module. The forward and inverse transformation of the IAT module implements the bijective mapping.
		This module takes the QLevel and feature as input to generate element-wise activation parameters $\beta$ and $\gamma$, further obtaining the output results.
		Thus, the forward and inverse procedures are mathematically invertible, enhancing the fidelity of reconstructed images.
	}
	\Description{IATM processing}
	\label{IATM}
\end{figure}

The forward transform of the IAT module is illustrated by pink arrows on the top of Figure~\ref{IATM}. The inputs are the quality level $L$ and the feature $ \textit{\textbf{s}} $. The element-wise activation parameters $\gamma \in \mathbb{R}^{c \times h \times w}$ and $\beta \in \mathbb{R}^{c \times h \times w}$ are then calculated by the IAT module from the quality level $L$ via convolutional operations. 
These activation parameters would be applied to the feature $\textit{\textbf{s}}$ via the Equation~\ref{equation_forward} to generate the feature $\textit{\textbf{e}}$,
\begin{equation}
	\textit{\textbf{e}} = (\textit{\textbf{s}} \odot \beta) \oplus \gamma,
	\label{equation_forward}
\end{equation}
where $\odot$ denotes the Hadamard product and $\oplus$ denotes the addition by element. $c$, $h$, and $w$ are the channel, height, and width of the feature, respectively. 
\par
The inverse transform of the IAT module is illustrated by green arrows at the bottom of Figure~\ref{IATM}. The input quality level $L$ and features $\hat{\textit{\textbf{e}}}$ are applied to obtain the feature $\hat{\textit{\textbf{s}}}$. 
This inverse transform is formulated by Equation~\ref{equation_inverse},
\begin{equation}
	\hat{\textit{\textbf{s}}} = (\hat{\textit{\textbf{e}}} \circleddash  \gamma) \oslash {\beta},
	\label{equation_inverse}
\end{equation}
where $\circleddash$ denotes the subtraction in elemental order, $\oslash$ denotes the division by elemental order.
%\par
Once the quality level $L$ is the same in both forward and inverse procedures, the invertibility of the operation between the features $\textit{\textbf{s}}$ and $\textit{\textbf{e}}$ can be guaranteed.
\par
In the previous work~\cite{chen2020variable}, a set of scaling factors was inserted directly before the quantizer to achieve the discrete adjustable compression rate. In our algorithm, the activation parameters are element-wise, which means that IAT module is computed as a spatial feature transform rather than a simple channel weighting. Moreover, the IAT module is attached after each invertible block which is initially proposed in RealNVP~\cite{dinh2016density} and adopted by baseline model~\cite{xie2021enhanced}, not just inserted before the quantizer. These adjustments not only make fine variable-rate adaptation available but also turn out to the better performance, the experiment "Impact of the QLevel Representation" in section~\ref{Impact_of_the_QLevel_Representation}  shows its effectiveness, and the results are shown in Figure~\ref{scale_factor}.

\subsection{Fine Variable-Rate Control}
Unlike interpolation-based methods~\cite{cui2021asymmetric, sun2021interpolation} for obtaining finer compression rates, our method achieves the fine compression rate adaptation directly by modulating the quality level $L$, which is more convenient when controlling the compression rate by only one parameter instead of two. Compared to Song et al.~\cite{song2021variable}, our method does not require additional semantic labels, either.
%Our method does not require additional semantic labels as well compared to Song et al. \cite{song2021variable}.
\par
The goal of lossy image compression is to minimize the length of the bits stream and the distortion between the source image $x$ and the reconstructed image $\hat{x}$. The optimization function is always expressed in the rate-distortion loss: $R+\lambda D$, where $\lambda $ is the Lagrange multiplier which determines the trade-off between the rate $R$ and the distortion $D$.
Theoretically, as long as the set of Lagrangian multiplier $\lambda$ is large enough, it is possible to achieve fine compression rate control, but in practice, the computational cost is too high.
For interpolation-based methods, the Lagrangian multiplier $\lambda$ is a scalar. Thus, at each iteration during training, only one element in a finite set of $\lambda$ would be randomly selected for optimization. 
In order to further promote the R-D performance of our model, we use a tensor instead of the scalar $ \lambda$.
Our optimization function implements fine variable-rate control by minimizing the rate-distortion loss $R+\Lambda \odot \textbf{D}$, where dimensions of $\Lambda \in \mathbb{R}^{C\times H\times W}$ and the distortion $\textbf{D} \in \mathbb{R}^{C\times H\times W}$ are the same as the dimension of the original input image. 
$\odot$ denotes the Hadamard product. 
In this formulation, $ \Lambda $ is a tensor and no longer a finite set of constant scalars. Thus, $\textbf{D}$ measures pixel-wise distortion and is defined as $\textbf{D}=\frac{\sum_{i=1}^{T}\lambda_{i}(x_{i}-\hat{x}_{i})^2}{T}$ , $T$ indicates the number of image pixels, $\lambda_{i}$ is the Lagrangian multiplier, $x_{i}$ and $\hat{x}_{i}$ denote one pixel of the orignal and reconstructed image, respectively. 
\par
$\Lambda$ is simply calculated from the quality level L via a monotonically increasing function:
$\Lambda=V(L)$, where $V:\mathbb{R}^{N} \to \mathbb{R}^{T}$. $V(L)=\theta \times e^{\tau \times L}$, $\theta=0.0012$, $\tau =4.382$, the process of dimensioning from $\mathbb{R}^{N}\rightarrow \mathbb{R}^{T}$ is done by direct replication between channels. $L=[l_{i}]_{i=1:N}$, $l_{i}\in [0, 1]$, $N=H\times W$, $T= C \times H\times W$. $C$, $H$, and $W$ denote the channel, height, and width of the source image $x$, respectively.
Under such a paradigm, we implement this pixel-wise distortion constraint by randomly generating values of each element of the tensor $ \Lambda $ via the quality level $L$ during training.
This is equivalent to increasing the number of $\lambda$ values selected at each iteration. So, the fine variable-rate control can be obtained by feeding exact quality levels during the testing.
\par
As in other learning-based method~\cite{balle2018variational}, the log-likelihood of the coded features $\hat{y}$ is estimated by a probabilistic model to replace the true compression rate $R$. Finally, the training loss would be:
\begin{equation}
	Loss=-log_{2}P_{\hat{y}}(\hat{y}|\Lambda)-log_{2}P_{\hat{z}}(\hat{z}|\Lambda)+\frac{\sum_{i=1}^{T}\lambda_{i}(x_{i}-\hat{x}_{i})^2}{T},
	\label{Loss_function}
\end{equation}
where $\hat{y}$ and $\hat{z}$ are quantized latent representations and side information respectively. $p_{\hat{y}}(\hat{y}|\Lambda)= \mathcal{N}(\mu,\sigma^{2})$, $\mu$ and $\sigma$ denote the estimates of the mean and standard deviation of the quantified latent representation $\hat{y}$. $p_{\hat{z}}(\hat{z}|\Lambda)=\mathcal{N}(\mu_{1},\sigma_{1}^{2})$, $\mu_{1}$ and $\sigma_{1}$ denote the estimates of the mean and standard deviation of the quantified side information $\hat{z}$. The side information usually represents the hyperprior originally proposed in \cite{balle2018variational} and refers to the extra stream $\hat{z}$ generated by the "Context Entropy Model" in Figure~\ref{Framework_architecture}.
It is worth noting that this loss function would be degraded to the standard rate-distortion optimization function if all elements of the tensor quality level $L$ are the same.
\par
In addition, our method can be trained on arbitrary unlabeled data instead of requiring semantic segmentation labels corresponding to the original data, which is different from Song et al.~\cite{song2021variable}, for training the model.

\begin{figure*}[!]
	\centering
	\subcaptionbox{PSNR on Kodak}{\includegraphics[width = 0.32\linewidth, height=4.6cm]{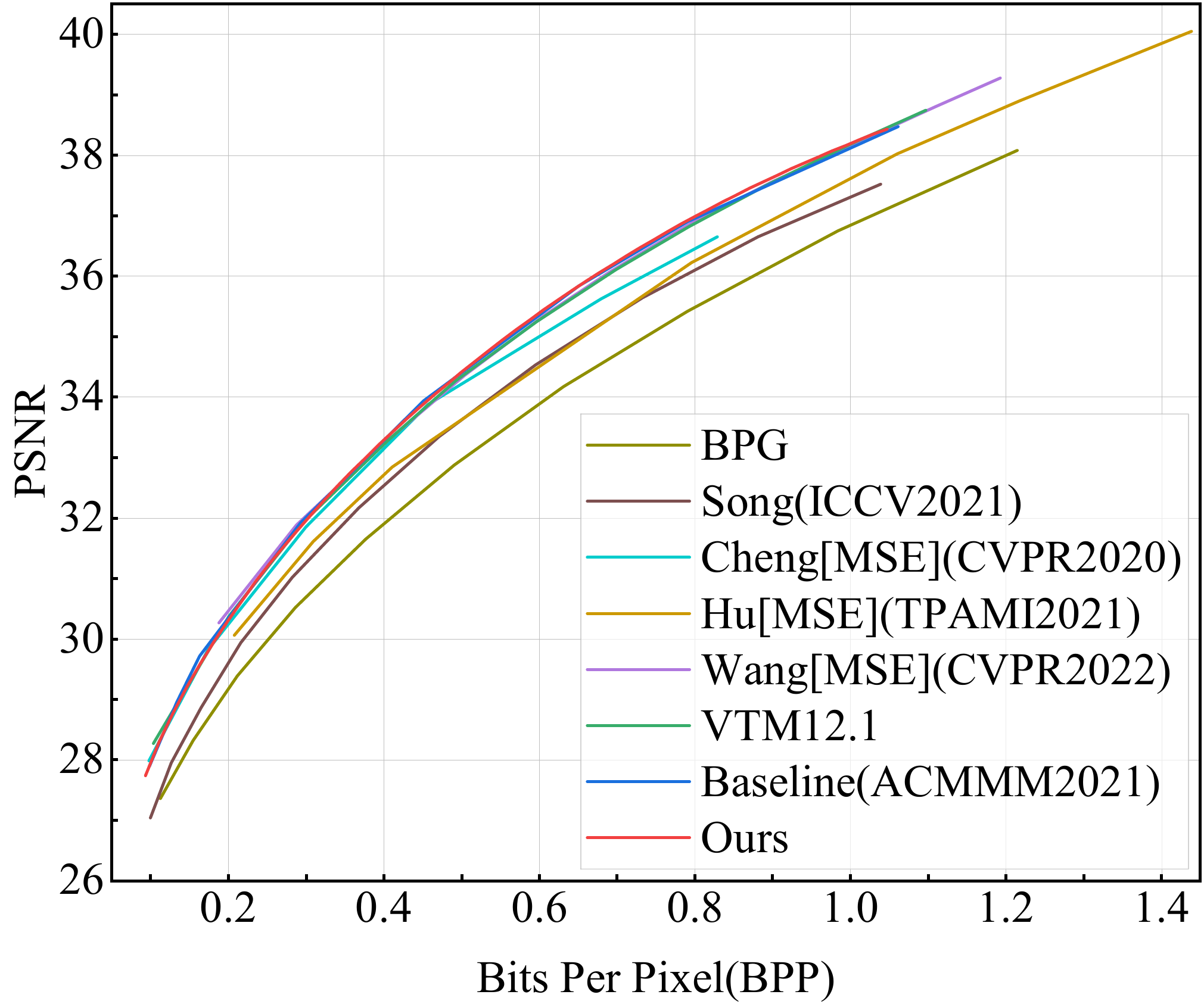}}\hfill
	\subcaptionbox{PSNR on CLIC} {\includegraphics[width = 0.32\linewidth, height=4.6cm]{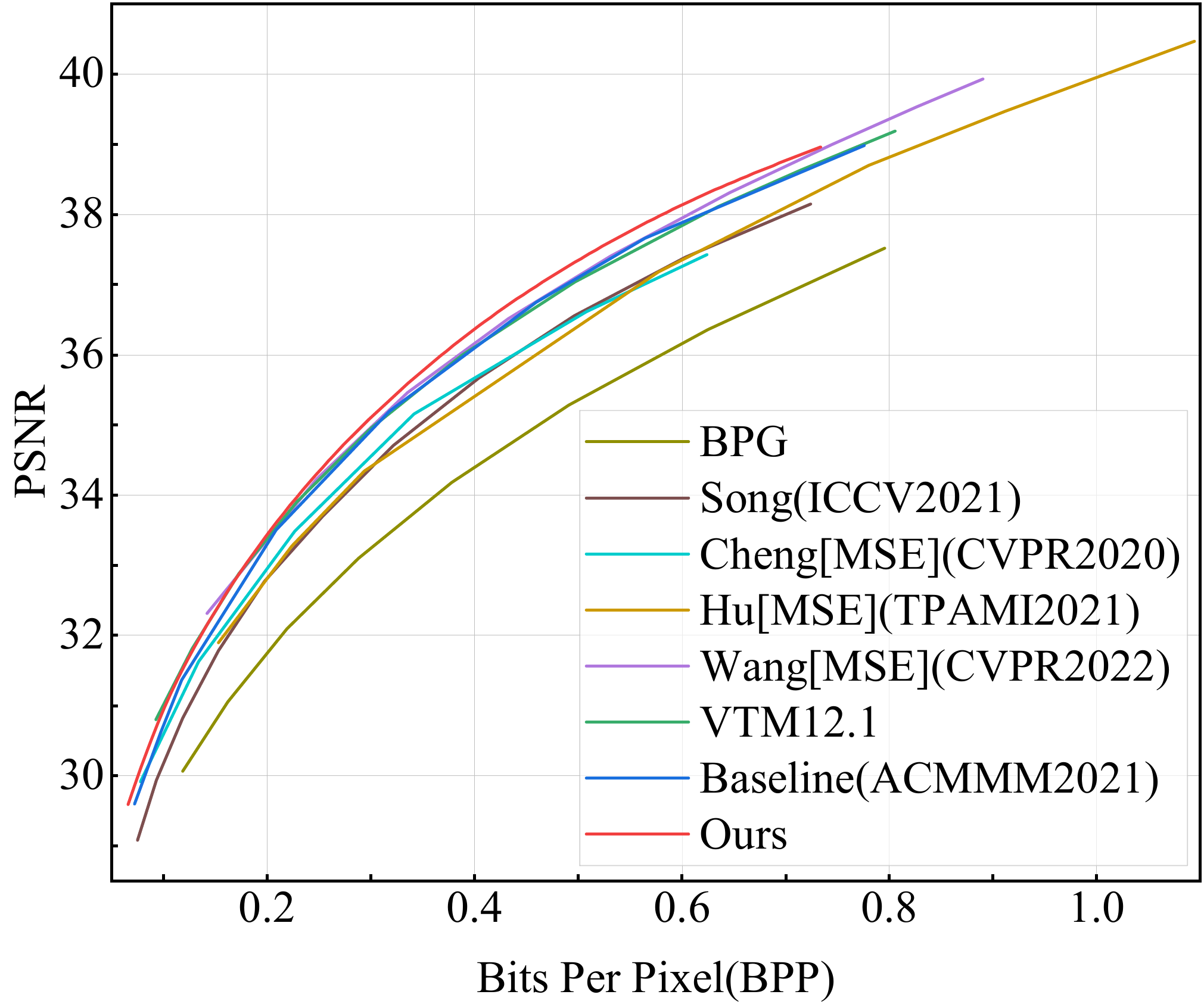}}\hfill
	\subcaptionbox{PSNR on DIV2K}{\includegraphics[width =0.32\linewidth, height=4.6cm]{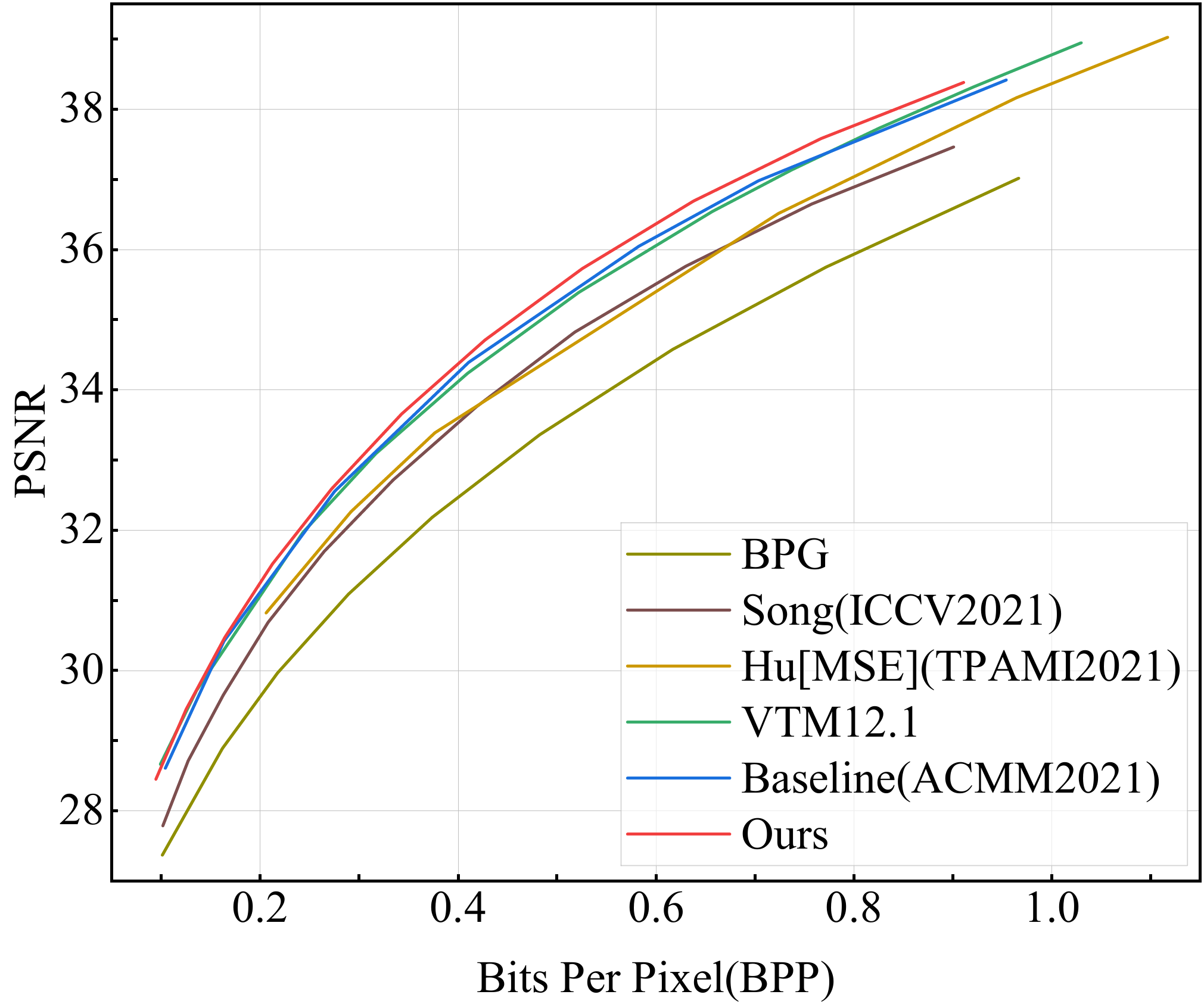}}\\[2ex]
	
	\subcaptionbox{MS-SSIM on Kodak}{\includegraphics[width = 0.32\linewidth, height=4.6cm]{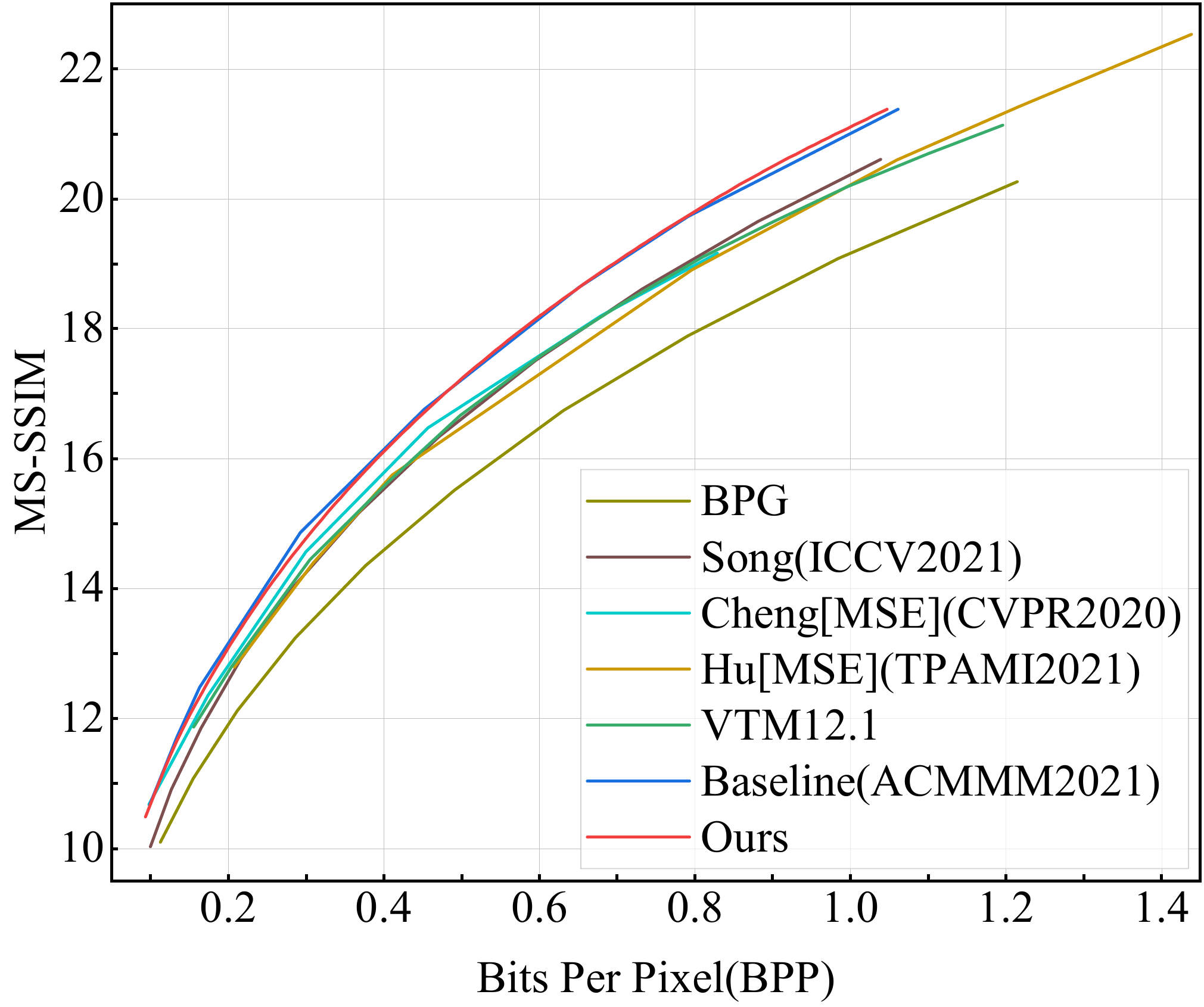}}\hfill
	\subcaptionbox{MS-SSIM on CLIC}{\includegraphics[width = 0.32\linewidth, height=4.6cm]{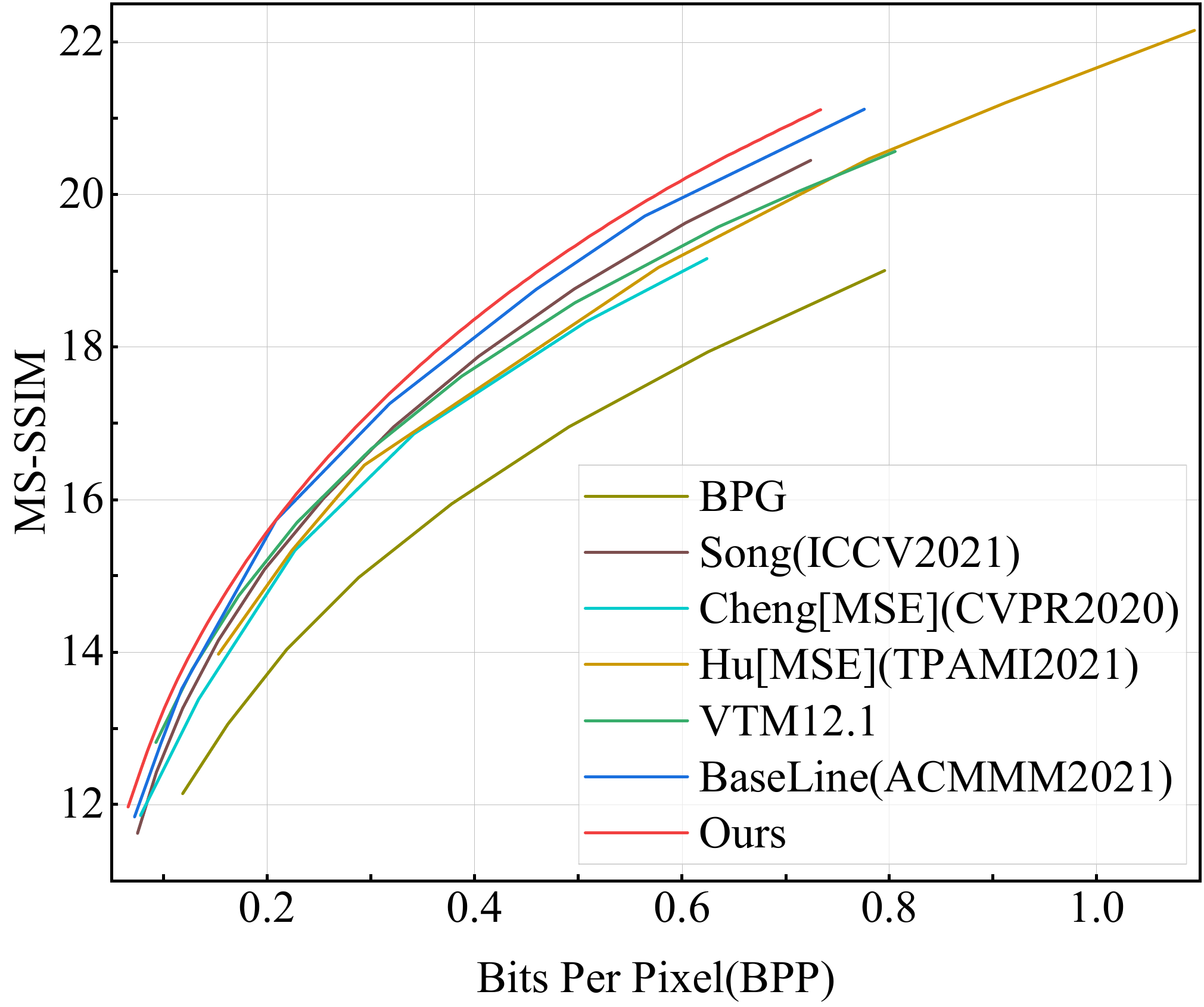}}\hfill
	\subcaptionbox{MS-SSIM on DIV2K}{\includegraphics[width = 0.32\linewidth, height=4.6cm]{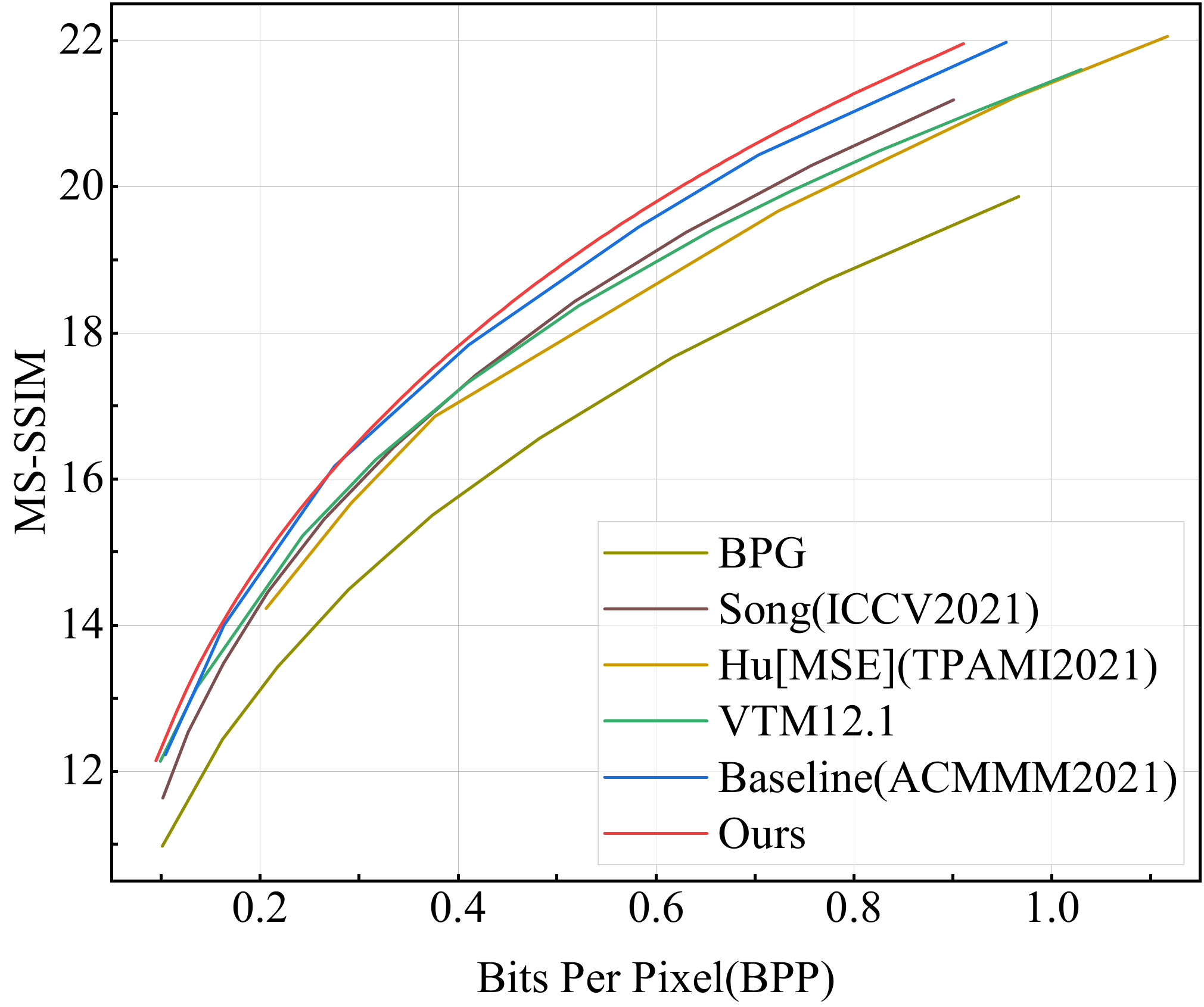}}
	\caption{RD performance curves aggregated over the Kodak, CLIC professional validation dataset, and DIV2K validation dataset. MS-SSIM values converted to decibels $(-10log_{10}(1-MS-SSIM))$. (a)/(b)/(c) and (d)/(e)/(f) are results on Kodak, CLIC, and DIV2K about PSNR and MS-SSIM, respectively. It is worth noting that CLIC and DIV2K are datasets with high-resolution images. That is, our method is especially effective on high-resolution images.}
	\label{performance_curve}
\end{figure*}

%\begin{figure*}[!]
%	\centering
%	\includegraphics[width = 0.96\linewidth]{figures/BPG_VTM_view_V3.pdf}
%	\caption{Visualization of sample images (ales-krivec-15949.png from CLIC dataset) reconstructed by BPG, VTM12.1, and Ours. The quantizer parameters (QPs) are used in BPG and VTM 12.1 to realize variable rate control. By adjusting the quality level (QL), our proposed method matches the rates of BPG and VTM 12.1 while outperforming them.}
%	\label{compare_with_classical_variable_rate}
%\end{figure*}

\section{Experiments}
\subsection{Implementation Details}
\subsubsection*{\textbf{Details For Training}}
In our implementation, the network of Xie et al.~\cite{xie2021enhanced} is adopted as our basic architecture.
The training datasets contain Flicker 2W~\cite{liu2020unified} and COCO~\cite{lin2014microsoft}. Flicker is used to train the network which has the context model, COCO is used to train the network without the context model. 
Our network is trained on $256 \times 256$ randomly cropped patches and discards images less than 256px in height or width during data pre-processing.
In training, the quality level $L$ needs to be sent to the INN section as a condition during the forward and inverse transform. The quality level $L$ takes a uniform value tensor between (0,1) during the testing and is randomly sampled between (0,1) during the training.
Our implementation relies on Pytorch~\cite{paszke2019pytorch} and an open-source CompressAI PyTorch library~\cite{begaint2020compressai}.
All experiments were conducted on RTX 3090 GPU and trained for about 2.5M iterations with batch size 8. Adam optimizer~\cite{Kingma2015Adam} is used to optimize the parameters, there were multistage learning rates $\{1e-4, 5e-5, 1e-5, 5e-6, 1e-6, 5e-7\}$ that changed with boundaries $\{1000000, 1300000, 1600000, 1900000, 2200000, 2500000\}$.
\subsubsection*{\textbf{Details For Testing}}
We evaluate the rate-distortion performance on three commonly used datasets. The Kodak~\cite{kodak} contains 24 lossless images with a size of $ 768 \times 512$. The CLIC Professional Validation dataset~\cite{CLIC2020} comprises 41 high-quality images with much higher resolution. The DIV2K validation dataset~\cite{Agustsson_2017_CVPR_Workshops} contains 100 images with high resolutions of 2K.
We draw curves based on the rate-distortion performance to compare the coding efficiency of different methods. We also calculate the area under the rate-distortion curve to observe the performance difference more effectively.

\subsection{Rate-Distortion Performance}
To verify the validity of the proposed approach, we conduct rate-distortion (RD) performance experiments on three datasets, \textit{i.e.}, Kodak~\cite{kodak}, CLIC~\cite{CLIC2020}, and DIV2K~\cite{Agustsson_2017_CVPR_Workshops}. We compare our approach with five recent state-of-the-art learning-based image compression methods~\cite{cheng2020learned, hu2021learning, xie2021enhanced, song2021variable, wang2022neural} and two classical codec methods, BPG~\cite{bpg} and VCC~\cite{vvc}. 
The results of learning-based methods are collected from their official GitHub pages or their papers. The VCC approach is implemented by the official Test Model VTM 12.1 with the intra-profile configuration from the official GitHub page to test images. Both VVC and BPG software were configured with the YUV444 format to maximize compression performance.
\par
All comparable results are demonstrated in Figure~\ref{performance_curve}. 
It is seen that our approach achieves the best results with commonly used metrics PSNR and MS-SSIM on three datasets. Compared with the baseline method~\cite{xie2021enhanced}, our approach achieves comparable R-D performance on the Kodak dataset (Figure~\ref{performance_curve} (a)(d)) and outperforms the baseline on both the CLIC dataset (Figure~\ref{performance_curve} (b)(e)) and the DIV2K dataset (Figure~\ref{performance_curve} (c)(f)). 
This means that our approach achieves the variable-rate adaptation based on the single rate method~\cite{xie2021enhanced} without sacrificing any performance, verifying the effectiveness of the IAT module.
It is worth noting that the CLIC dataset and DIV2K dataset are high-resolution images, implying that our method is more effective on high-resolution images. Our approach empowers the network model with variable-rate in addition to improving the algorithmic performance of the original model.
To further compare the performance between the baseline~\cite{xie2021enhanced} and our method, we calculate their corresponding area under curve (AUC) values, as shown in Table~\ref{tab_AUC}. The results show that our approach outperforms the single rate model method by Xie et al.~\cite{xie2021enhanced} in terms of the aggregated AUC metric. 
\par
In addition, Our proposed method could achieve variable-rate image compression with a fine granularity. To verify the effectiveness of fine variable-rate control, we illustrate multiple performances of fine variable-rate control within the low and high bpp range in Table~\ref{fine_variable_rate_test}. In practice, classical image codecs provide hundreds of variable-rate RD points to meet the basic requirement of the application. Compared with that, our method obtains at least 1000 effective variable-rate RD points with a very fine PSNR and MS-SSIM. We achieved the fine-rate control compared with the classical image codecs BPG~\cite{bpg} and VTM 12.1~\cite{vvc}, the comparative results refer to the supplementary material.

\begin{table}[h]

	\caption{Area under curve (AUC) of our method and Xie et al.~\cite{xie2021enhanced}(Baseline) on different datasets about PSNR and MS-SSIM. The bpp range is determined by the intersection of two methods. Our approach makes a single-rate baseline compression model achieve the variable-rate ability and even outperforms the baseline in R-D performance.}

	\begin{center}
		\resizebox{\linewidth}{!}{
			\begin{tabular}{c|cc|cc}
				\hline
				\multirow{2}{*}{Dataset} & \multicolumn{2}{c|}{Xie et al. \cite{xie2021enhanced}} & \multicolumn{2}{c}{Ours} \\ 
				\cline{2-5}
				& $AUC_{PSNR}$   & $AUC_{MS-SSIM}$  & $AUC_{PSNR}$     & $AUC_{MS-SSIM}$ \\
				\hline
				Kodak   & 32.7866        &16.5030           & \textbf{32.7883} & \textbf{16.5036} \\
				CLIC    & 23.5896        &11.7463           & \textbf{23.7082} & \textbf{11.8571} \\
				DIV2K   & 28.0998        &14.7868           & \textbf{28.2138} & \textbf{14.8901} \\ 
				\hline
		\end{tabular}}
	\end{center}
	\label{tab_AUC}
\end{table}

\begin{table}[h]
	\centering
	\caption{Variable-rate control experiments over the Kodak dataset. Our approach can finely control the compression rate within the whole bpp range (no matter low or high). 
	}

	\label{fine_variable_rate_test}
%	 \scalebox{0.85}

	\resizebox{\linewidth}{!}{
		\begin{tabular}{ccc|ccc}
			\hline
			\multicolumn{3}{c|}{LOW}      & \multicolumn{3}{c}{HIGH} \\
			\hline
			BPP & PSNR(dB) & MS-SSIM(dB)  & BPP & PSNR(dB) & MS-SSIM(dB) \\
			\hline
			0.28181  & 31.6951  & 14.5015 & 1.02433 & 38.3226 & 21.2580 \\
			0.28265  & 31.7071  & 14.5153 & 1.02587 & 38.3312 & 21.2664 \\
			0.28342  & 31.7177  & 14.5263 & 1.02733 & 38.3388 & 21.2717 \\
			0.28416  & 31.7291  & 14.5377 & 1.02910 & 38.3468 & 21.2819 \\
			0.28500  & 31.7435  & 14.5517 & 1.03071 & 38.3548 & 21.2903 \\
			0.28576  & 31.7538  & 14.5639 & 1.03250 & 38.3625 & 21.2995 \\
			0.28659  & 31.7657  & 14.5765 & 1.03406 & 38.3703 & 21.3087 \\
			0.28734  & 31.7761  & 14.5874 & 1.03564 & 38.3767 & 21.3190 \\
			0.28808  & 31.7884  & 14.5952 & 1.03733 & 38.3872 & 21.3291 \\
			0.28880  & 31.8004  & 14.6092 & 1.03885 & 38.3943 & 21.3355 \\
			\hline
	\end{tabular}}

\end{table}

\begin{figure*}[!]
	\centering
	\includegraphics[width = 0.96\linewidth, height = 10cm]{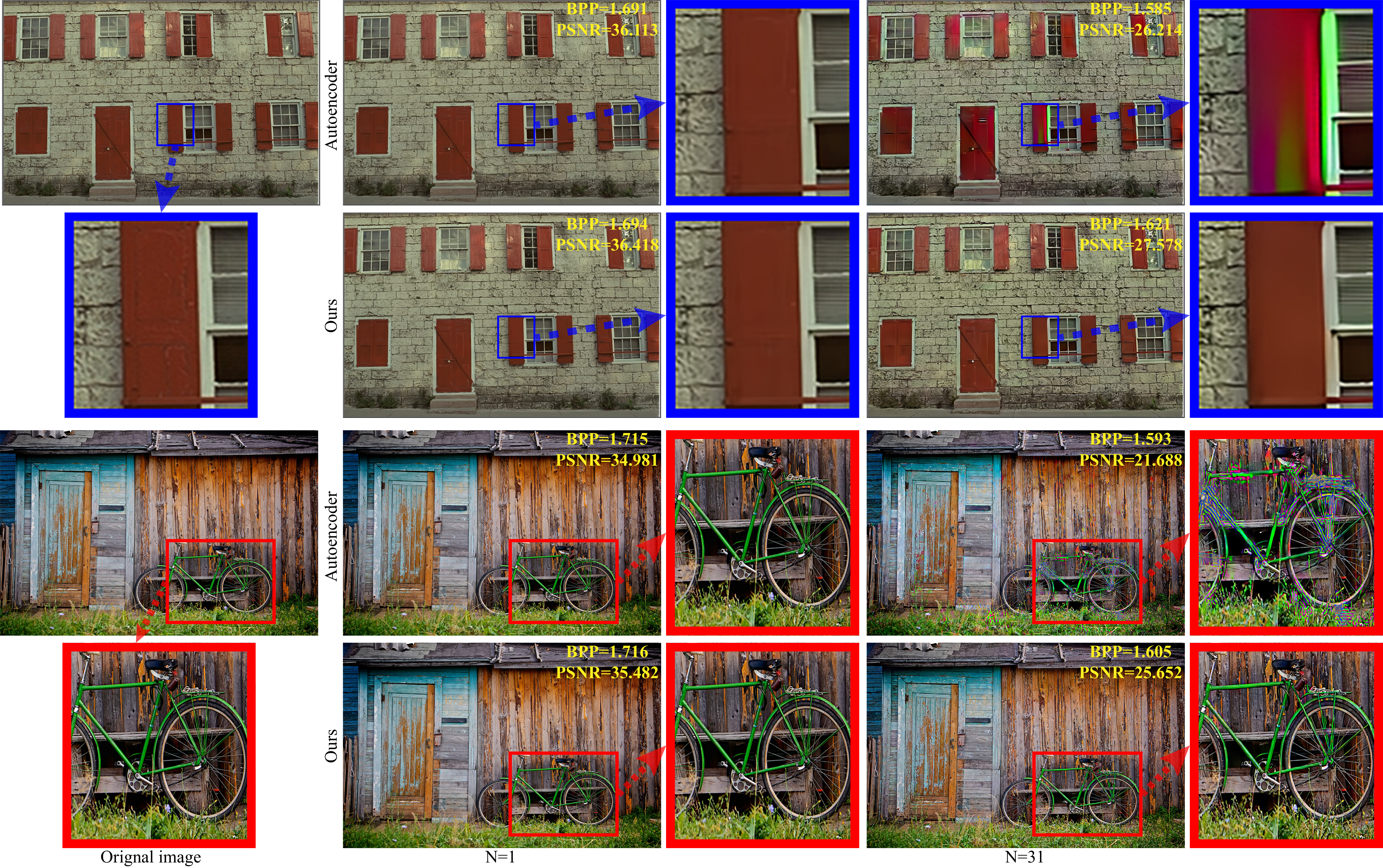}
	\caption{Qualitative results after different numbers of compression/decompression operations under various rates. The two images (kodim1.png and alexander-shustov-73.png) are from the Kodak dataset and the CLIC dataset, respectively. 
		Severe artifacts and color shifts would appear in the state-of-the-art VAE-based approach~\cite{song2021variable} once multiple operations are executed, in contrast to better fidelity maintenance of our approach.
		Please refer to the supplementary material for more cases. N indicates the number of compression/decompression operations. Best viewed in color.}
	% \vspace{-0.1cm}
	\label{result-re_ecoding}
\end{figure*}

%\subsection{Fine Variable-Rate Control}
%Our proposed method could achieve variable-rate image compression with a fine granularity. To verify the effectiveness of fine control, we illustrate multiple performances of fine rate control within the low and high bpp range in Table~\ref{fine_variable_rate_test}. In practice, classical image codecs provide hundreds of variable-rate RD points to meet the basic requirement of the application. Compared with that, our method obtains at least 1000 effective variable-rate RD points with a very fine PSNR and MS-SSIM. We achieved the fine-rate control compared with the classical image codecs and comparative results refer to the supplementary material.

%Figure~\ref{compare_with_classical_variable_rate} realizes a fine rate control compared with the classical image codecs. Figure~\ref{compare_with_classical_variable_rate} also shows that our proposed method provides better reconstruction quality on PSNR, MS-SSIM, with fewer artifacts in visual perception.

\subsection{Fidelity for Re-encoding}
In order to verify the high fidelity, our method is compared with the latest VAE-based variable-rate method by Song et al.~\cite{song2021variable}. This method does not use context model and has available source code.
To make a fair comparison, we remove the context model and add the non-local attention module~\cite{chen2021end} to the hyperprior layer.
Figure~\ref{result-re_ecoding} illustrates the results of multiple compression/decompression operations on the same image with different compression rates. With operations increasing, our proposed method shows higher fidelity.
\par
\begin{figure}[b]
	\subcaptionbox{Various rates on PSNR}{\includegraphics[width = 0.49\linewidth, height=3.8cm]{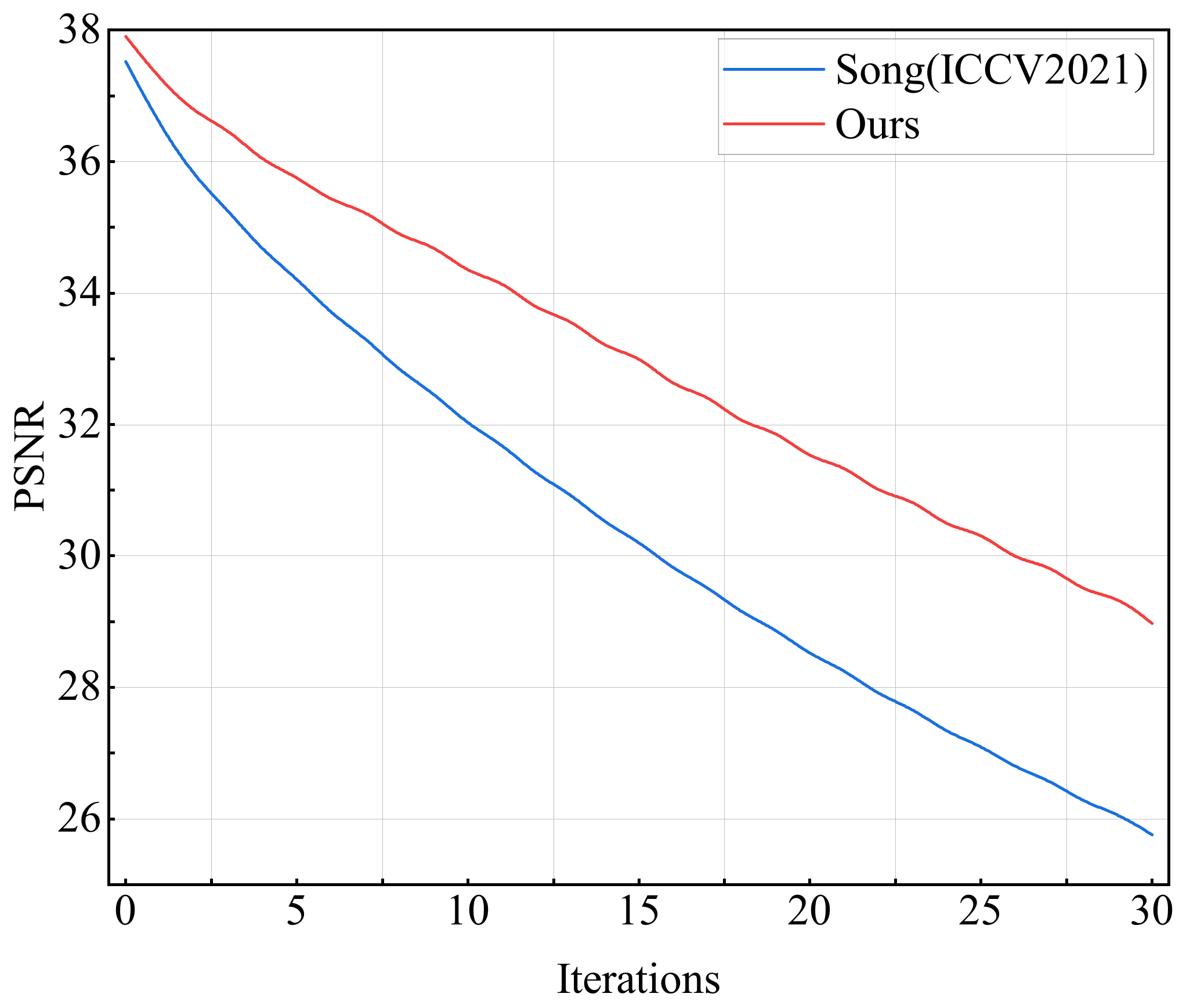}}\hfill
	\subcaptionbox{Fixed rate on PSNR}{\includegraphics[width = 0.49\linewidth, height=3.8cm]{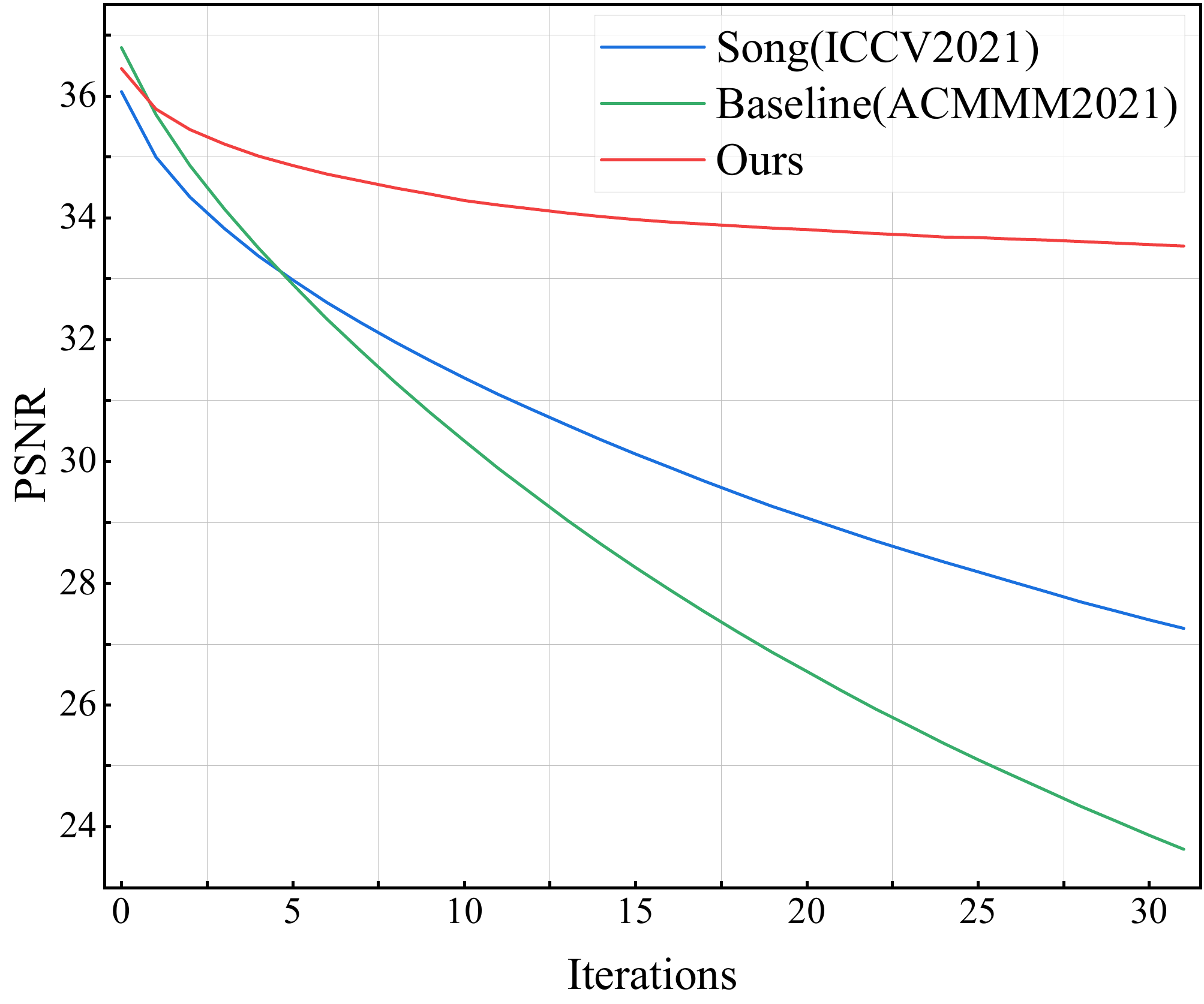}}\\[2ex]
	\subcaptionbox{Various rate on MS-SSIM}{\includegraphics[width = 0.49\linewidth, height=3.8cm]{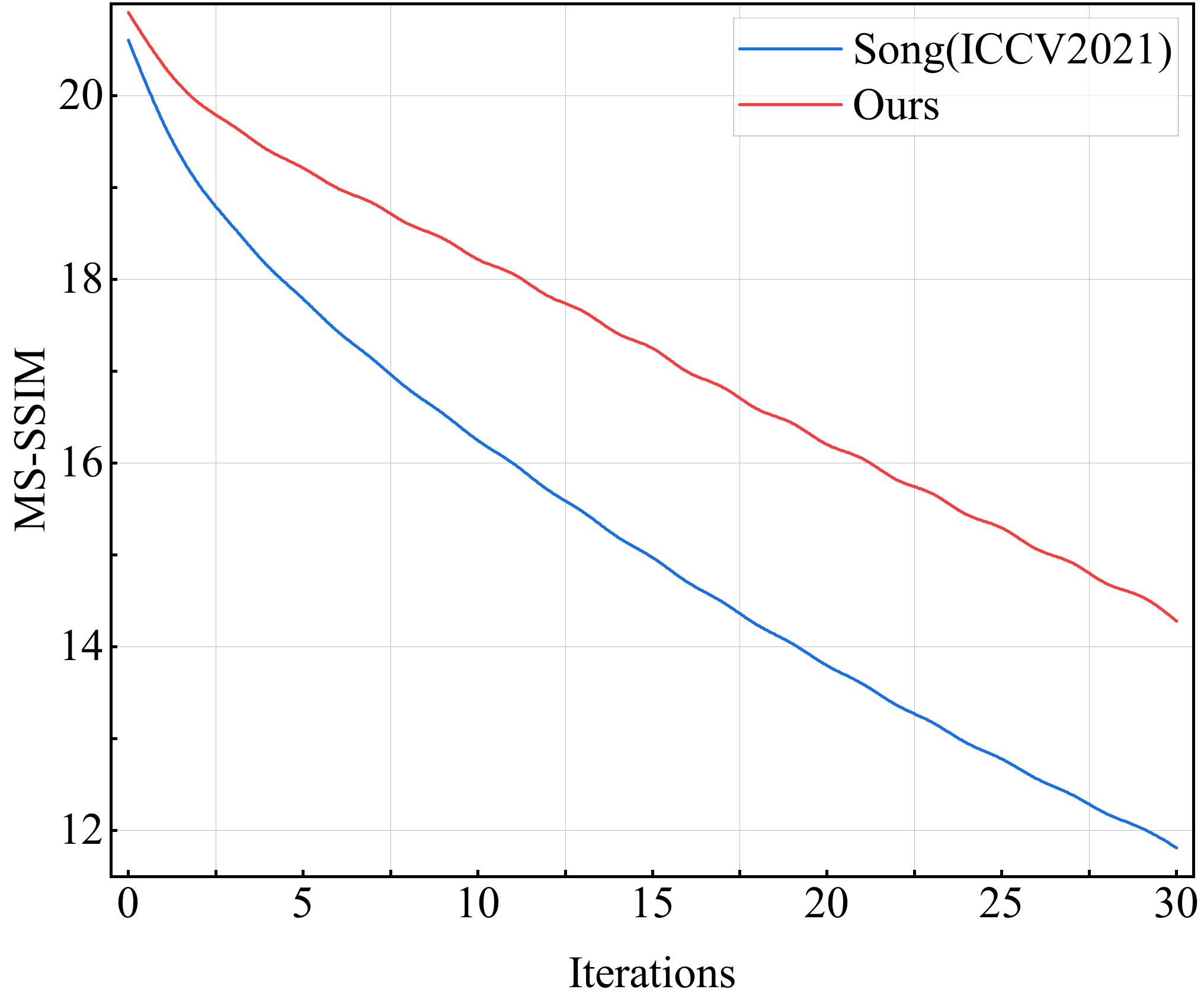}}\hfill
	\subcaptionbox{Fixed rate on MS-SSIM}{\includegraphics[width = 0.49\linewidth, height=3.8cm]{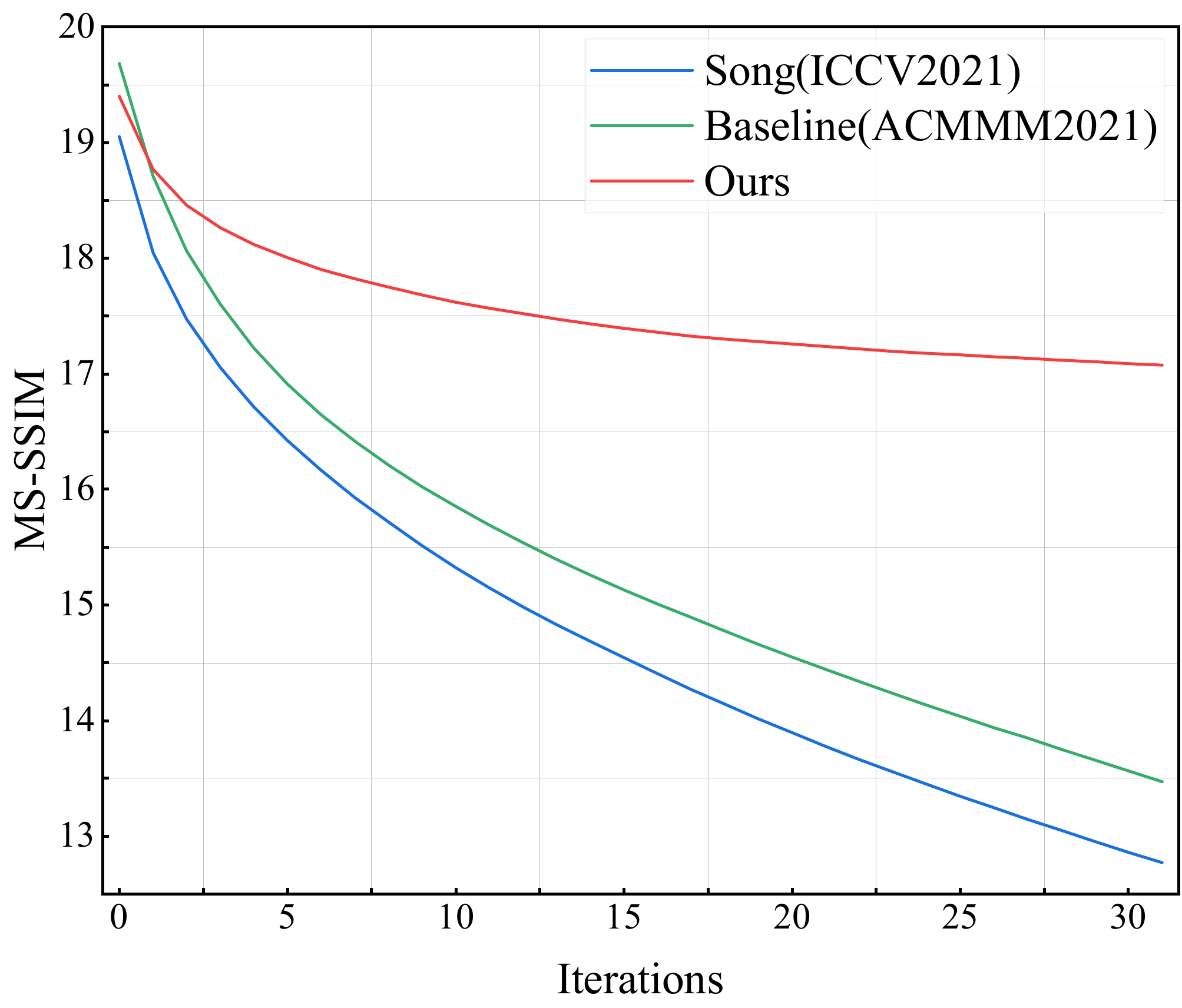}}
	\caption{Successive re-encodings on the Kodak dataset. (a) and (c): Compression rates of each compression/decompression operation are different. (b) and (d): The compression rate is fixed. Our approach outperforms baseline~\cite{xie2021enhanced} and Song et al.~\cite{song2021variable}~(a SOTA variable-rate approach) by a large margin to show the superiority of fidelity maintenance especially when multiple operations are executed.} 
	\label{Iterations_performance_curve}
\end{figure}

Figure~\ref{Iterations_performance_curve} (a)(c) show the rate-distortion performance after multiple operations of compression/decompression with different compression rates. Both approaches change from high to low bpp ranges, our method in the set of bpp \{ 1.0267, 1.0116, 0.9949, 0.9784, 0.9619, 0.9456, 0.9292, 0.9127, 0.8965, 0.8809, 0.8658, 0.8507, 0.8357, 0.8206, 0.8056, 0.7907 \}, Song et al.~\cite{song2021variable} in the set of bpp \{1.0392, 1.0249, 1.0091, 0.9932, 0.9768, 0.9606, 0.9449, 0.9287, 0.9128, 0.8968, 0.8813, 0.8658, 0.8505, 0.8351, 0.8201, 0.8052\}. It is clearly seen that our method outperforms Song et al.~\cite{song2021variable}, after multiple compression/decompression operations. 
Figure~\ref{Iterations_performance_curve} (b)(d) show the rate-distortion performance by multiple operations with the fixed compression rate. Both approaches achieve a bit rate of 0.791 bpp for all steps. Also, our method achieves better results significantly, compared with Song et al.~\cite{song2021variable} and baseline~\cite{xie2021enhanced}. The results indicate that our proposed IAT module is powerful to maintain image fidelity, which is important for practical applications.
It is noteworthy that compression methods capable of high-fidelity in re-encoding are of great importance in the video production pipeline, as image/video content may be edited/composited by different people or at different times, requiring re-encodings in the process.

\subsection{Discussion}
\subsubsection*{\textbf{Impact of the QLevel Representation}}
\label{Impact_of_the_QLevel_Representation}
\begin{figure}[h]
	\centering
	\subcaptionbox{PSNR on Kodak}{\includegraphics[width = 0.49\linewidth, height=3.8cm]{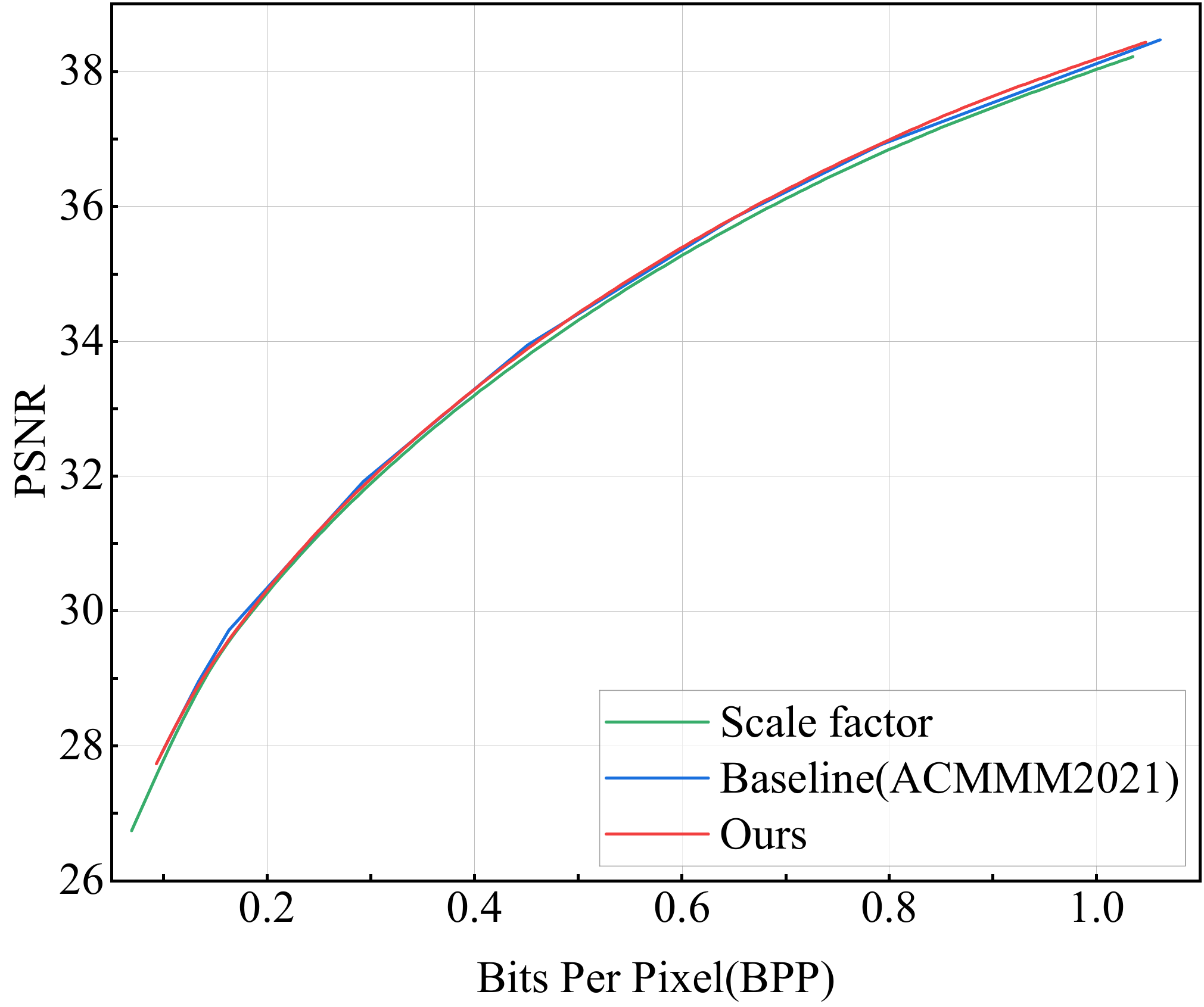}}\hfill
	\subcaptionbox{PSNR on CLIC}{\includegraphics[width = 0.49\linewidth, height=3.8cm]{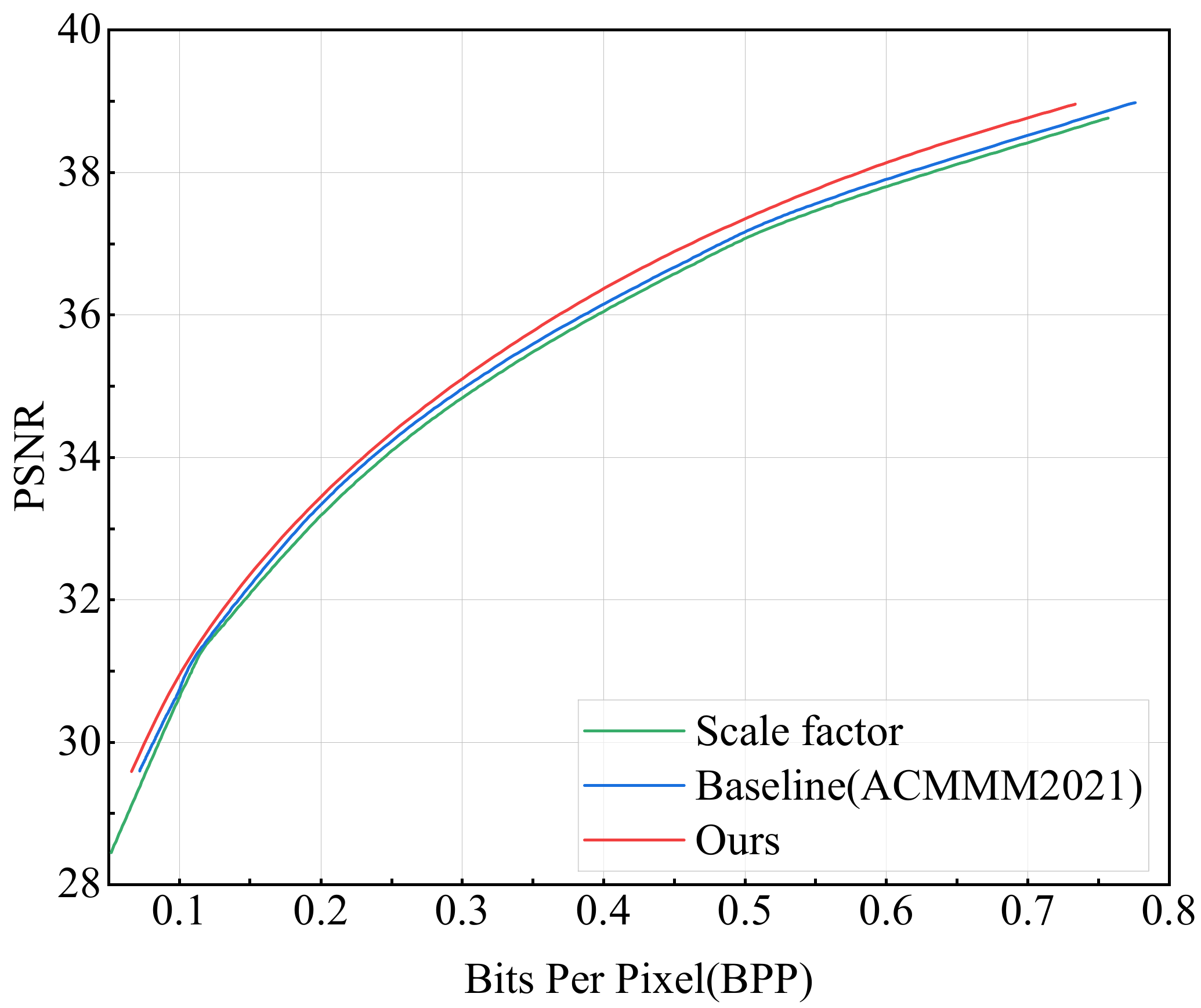}}
	\caption{Impact of the QLevel Representation. The scale factor method (green line) is similar to Chen et al. \cite{chen2020variable}.
		Our proposed tensor-based QLevel representation achieves better performance than simply using a scalar to control the compression rate.}
	\label{scale_factor}
\end{figure}

To further analyze the effectiveness of the tensor-based QLevel representation of our IAT module, we conducted an ablation study by modifying the quality level representation. We compared our approach with the baseline method~\cite{xie2021enhanced} and the simplified version of our method, which modifies the quality level from tensor to scalar, similar to~\cite{chen2020variable}. Comparative results are shown in Figure~\ref{scale_factor}. The results indicate that the proposed tensor-based quality level obtains better performance, compared with the scalar factor ones, which only provides channel-wise weighted computations on latent representation.

\subsubsection*{\textbf{Impact of Gain Components}}
\begin{figure}[h]
	\includegraphics[width = 0.5\linewidth, height=3.6cm]{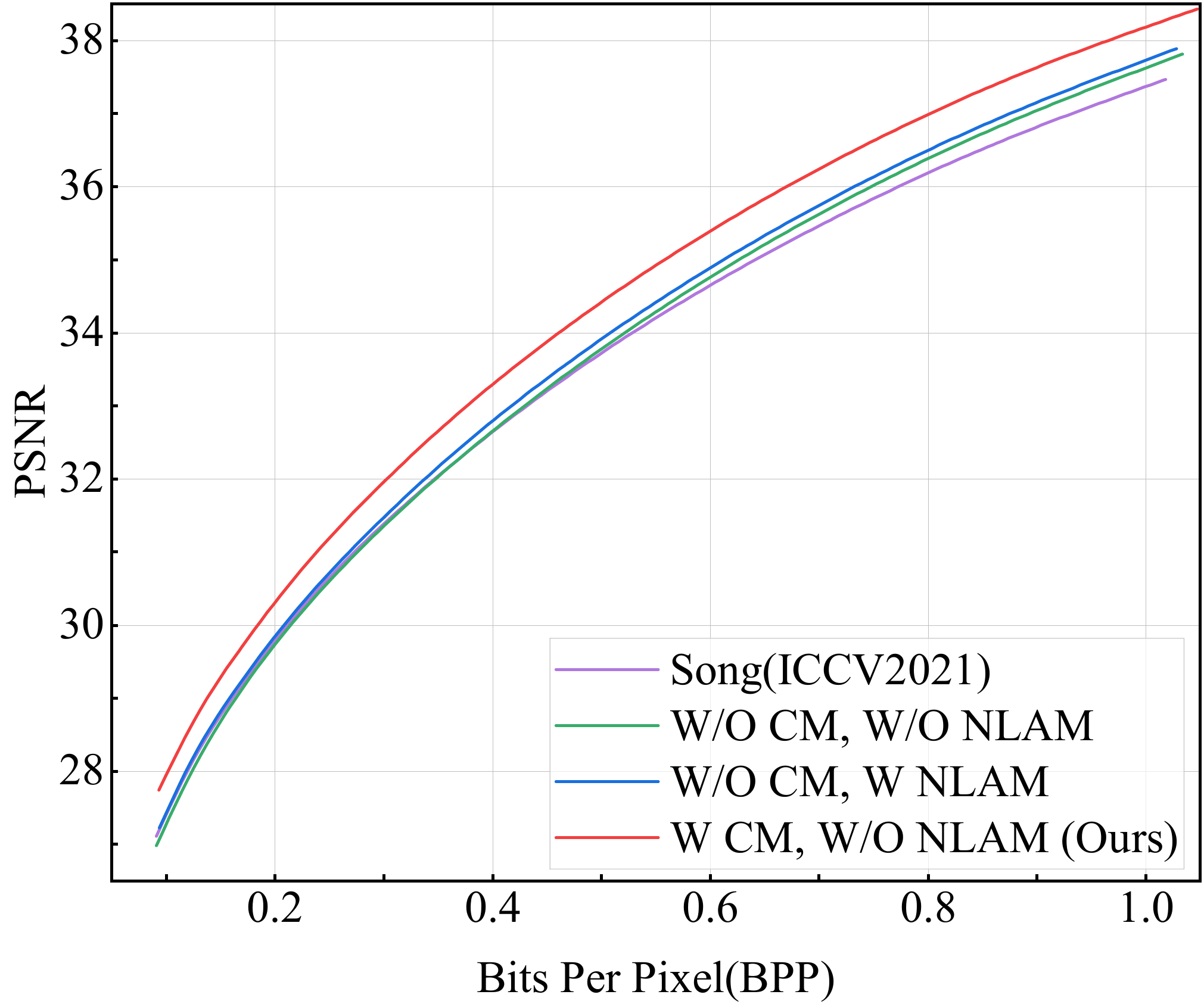}
	\caption{Impact of Gain Components. W/O represents `without', W represents `with', CM represents `context model', and NLAM represents `non-local attention module'.}
	\label{impact_of_non_local_attention}
\end{figure}

The context model~\cite{lee2018context, mentzer2018conditional, minnen2018joint} and the non-local attention module~\cite{chen2021end} are commonly used in the learned-based image compression methods to further reduce statistical redundancy within the latent features and improve the probabilistic estimation ability of the network. We conduct an ablation study to evaluate the impact of the context model and non-local attention module on our method in the Kodak dataset, as shown in Figure~\ref{impact_of_non_local_attention}. We start from a baseline without the context model and non-local attention module, \textit{i.e.}, W/O CM (context model) and W/O NLAM (non-local attention module), and plot the rate-distortion performance in green color. Then, we add the non-local attention module (blue color) and context model (red color) to evaluate the performance. We can observe that using the context model achieves the best results, while it requires high computational costs~(codec process takes about 233 seconds on an Intel(R) Core(TM) i9-10900K CPU). In addition, Our method outperforms Song et al.~\cite{song2021variable} without the context model and non-local attention module, demonstrating the effectiveness of the proposed method. 
%\subsubsection*{\textbf{Impact of  the IAT Module in Re-encodings}}
%To further analyze the effectiveness of the fidelity maintenance property between ours and baseline, Ours and the baseline method conducted fixed compression rate re-encoding experiments (bpp=0.318) on the CLIC Professional Validation dataset~\cite{CLIC2020}, respectively. The results indicate that our proposed IAT module is powerful to maintain image fidelity, which is important for practical applications.
%

%\begin{figure}
%	\subcaptionbox{Fixed rate on PSNR}{\includegraphics[width = 0.49\linewidth]{figures/Fix_bpp_PSNR_baseline_ours.pdf}}\hfill
%	\subcaptionbox{Fixed rate on MS-SSIM}{\includegraphics[width = 0.49\linewidth]{figures/Fix_bpp_MS_SSIM_baseline_ours.pdf}}
%
%	\caption{Successive re-encodings on the CLIC Professional Validation dataset~\cite{CLIC2020}. Our approach outperforms Baseline~\cite{xie2021enhanced} and Song et al.~\cite{song2021variable}~(a SOTA variable-rate approach) by a large margin to show the superiority of fidelity maintenance especially when multiple operations are executed.} 
%	\label{Iterations_performance_curve_basline_ours}
%\end{figure}

\section{Conclusion}
In this paper, we propose a high-fidelity variable-rate image compression method by introducing the Invertible Activation Transformation~(IAT) module. 
The IAT module, implemented in a mathematical invertible manner, as a feature activation transform layer of the invertible neural network, has the ability of fine variable-rate control by feeding the quality level~(QLevel) to generate the scaling and bias tensors while better maintaining the image fidelity.
Extensive experiments demonstrate that the single rate model equipped with our IAT module is able to achieve fine variable-rate control without any performance compromise.
Thanks to the mathematical invertibility of our approach, fewer artifacts or color shifts would have appeared and the fidelity of reconstructed images is better maintained, especially when multiple re-encodings are executed under various compression rates. 

%\section*{Acknowledgement}
\begin{acks}
%This work was supported by the National Key Laboratory Foundation of China grant No. 6142113200307.
This work was supported by the National Natural Science Foundation of China (NSFC) grant No. 62176100, the Special Project of Science and Technology Development of Central guiding LocalCentral Guidance on Local Science and Technology Development Fund of Hubei Province grant 2021BEE056 and the National Key Laboratory Foundation of China grant No. 6142113200307.
\end{acks}

\clearpage
\newpage
\bibliographystyle{ACM-Reference-Format}
\balance
\bibliography{reference}

%%% -*-BibTeX-*-
%%% Do NOT edit. File created by BibTeX with style
%%% ACM-Reference-Format-Journals [18-Jan-2012].

\begin{thebibliography}{57}

%%% ====================================================================
%%% NOTE TO THE USER: you can override these defaults by providing
%%% customized versions of any of these macros before the \bibliography
%%% command.  Each of them MUST provide its own final punctuation,
%%% except for \shownote{}, \showDOI{}, and \showURL{}.  The latter two
%%% do not use final punctuation, in order to avoid confusing it with
%%% the Web address.
%%%
%%% To suppress output of a particular field, define its macro to expand
%%% to an empty string, or better, \unskip, like this:
%%%
%%% \newcommand{\showDOI}[1]{\unskip}   % LaTeX syntax
%%%
%%% \def \showDOI #1{\unskip}           % plain TeX syntax
%%%
%%% ====================================================================

\ifx \showCODEN    \undefined \def \showCODEN     #1{\unskip}     \fi
\ifx \showDOI      \undefined \def \showDOI       #1{#1}\fi
\ifx \showISBNx    \undefined \def \showISBNx     #1{\unskip}     \fi
\ifx \showISBNxiii \undefined \def \showISBNxiii  #1{\unskip}     \fi
\ifx \showISSN     \undefined \def \showISSN      #1{\unskip}     \fi
\ifx \showLCCN     \undefined \def \showLCCN      #1{\unskip}     \fi
\ifx \shownote     \undefined \def \shownote      #1{#1}          \fi
\ifx \showarticletitle \undefined \def \showarticletitle #1{#1}   \fi
\ifx \showURL      \undefined \def \showURL       {\relax}        \fi
% The following commands are used for tagged output and should be
% invisible to TeX
\providecommand\bibfield[2]{#2}
\providecommand\bibinfo[2]{#2}
\providecommand\natexlab[1]{#1}
\providecommand\showeprint[2][]{arXiv:#2}

\bibitem[Agustsson and Timofte(2017)]%
        {Agustsson_2017_CVPR_Workshops}
\bibfield{author}{\bibinfo{person}{Eirikur Agustsson} {and}
  \bibinfo{person}{Radu Timofte}.} \bibinfo{year}{2017}\natexlab{}.
\newblock \showarticletitle{NTIRE 2017 Challenge on Single Image
  Super-Resolution: Dataset and Study}. In \bibinfo{booktitle}{\emph{CVPRW}}.
\newblock


\bibitem[Agustsson et~al\mbox{.}(2019)]%
        {agustsson2019generative}
\bibfield{author}{\bibinfo{person}{Eirikur Agustsson}, \bibinfo{person}{Michael
  Tschannen}, \bibinfo{person}{Fabian Mentzer}, \bibinfo{person}{Radu Timofte},
  {and} \bibinfo{person}{Luc~Van Gool}.} \bibinfo{year}{2019}\natexlab{}.
\newblock \showarticletitle{Generative Adversarial Networks for Extreme Learned
  Image Compression}. In \bibinfo{booktitle}{\emph{ICCV}}.
\newblock


\bibitem[Ardizzone et~al\mbox{.}(2019)]%
        {ardizzone2018analyzing}
\bibfield{author}{\bibinfo{person}{Lynton Ardizzone}, \bibinfo{person}{Jakob
  Kruse}, \bibinfo{person}{Sebastian Wirkert}, \bibinfo{person}{Daniel Rahner},
  \bibinfo{person}{Eric~W Pellegrini}, \bibinfo{person}{Ralf~S Klessen},
  \bibinfo{person}{Lena Maier-Hein}, \bibinfo{person}{Carsten Rother}, {and}
  \bibinfo{person}{Ullrich K{\"o}the}.} \bibinfo{year}{2019}\natexlab{}.
\newblock \showarticletitle{Analyzing Inverse Problems with Invertible Neural
  Networks}. In \bibinfo{booktitle}{\emph{ICLR}}.
\newblock


\bibitem[Ball{\'e} et~al\mbox{.}(2017)]%
        {balle2016end}
\bibfield{author}{\bibinfo{person}{Johannes Ball{\'e}}, \bibinfo{person}{Valero
  Laparra}, {and} \bibinfo{person}{Eero~P Simoncelli}.}
  \bibinfo{year}{2017}\natexlab{}.
\newblock \showarticletitle{End-to-End Optimized Image Compression}. In
  \bibinfo{booktitle}{\emph{ICLR}}.
\newblock


\bibitem[Ball{\'e} et~al\mbox{.}(2018)]%
        {balle2018variational}
\bibfield{author}{\bibinfo{person}{Johannes Ball{\'e}}, \bibinfo{person}{David
  Minnen}, \bibinfo{person}{Saurabh Singh}, \bibinfo{person}{Sung~Jin Hwang},
  {and} \bibinfo{person}{Nick Johnston}.} \bibinfo{year}{2018}\natexlab{}.
\newblock \showarticletitle{Variational Image Compression with a Scale
  Hyperprior}. In \bibinfo{booktitle}{\emph{ICLR}}.
\newblock


\bibitem[B{\'e}gaint et~al\mbox{.}(2020)]%
        {begaint2020compressai}
\bibfield{author}{\bibinfo{person}{Jean B{\'e}gaint}, \bibinfo{person}{Fabien
  Racap{\'e}}, \bibinfo{person}{Simon Feltman}, {and} \bibinfo{person}{Akshay
  Pushparaja}.} \bibinfo{year}{2020}\natexlab{}.
\newblock \showarticletitle{CompressAI: a Pytorch Library and Evaluation
  Platform for End-to-End Compression Research}.
\newblock \bibinfo{journal}{\emph{arXiv preprint arXiv:2011.03029}}
  (\bibinfo{year}{2020}).
\newblock


\bibitem[Bellard(2015)]%
        {bpg}
\bibfield{author}{\bibinfo{person}{Fabrice Bellard}.}
  \bibinfo{year}{2015}\natexlab{}.
\newblock \bibinfo{title}{BPG Image Format}.
\newblock
\newblock
\urldef\tempurl%
\url{https://bellard.org/bpg/}
\showURL{%
\tempurl}


\bibitem[Chen et~al\mbox{.}(2021)]%
        {chen2021end}
\bibfield{author}{\bibinfo{person}{Tong Chen}, \bibinfo{person}{Haojie Liu},
  \bibinfo{person}{Zhan Ma}, \bibinfo{person}{Qiu Shen}, \bibinfo{person}{Xun
  Cao}, {and} \bibinfo{person}{Yao Wang}.} \bibinfo{year}{2021}\natexlab{}.
\newblock \showarticletitle{End-to-End Learnt Image Compression via Non-Local
  Attention Optimization and Improved Context Modeling}.
\newblock \bibinfo{journal}{\emph{IEEE Transactions on Image Processing}}
  \bibinfo{volume}{30} (\bibinfo{year}{2021}), \bibinfo{pages}{3179--3191}.
\newblock


\bibitem[Chen and Ma(2020)]%
        {chen2020variable}
\bibfield{author}{\bibinfo{person}{Tong Chen} {and} \bibinfo{person}{Zhan Ma}.}
  \bibinfo{year}{2020}\natexlab{}.
\newblock \showarticletitle{Variable Bitrate Image Compression with Quality
  Scaling Factors}. In \bibinfo{booktitle}{\emph{ICASSP}}.
\newblock


\bibitem[Cheng et~al\mbox{.}(2020)]%
        {cheng2020learned}
\bibfield{author}{\bibinfo{person}{Zhengxue Cheng}, \bibinfo{person}{Heming
  Sun}, \bibinfo{person}{Masaru Takeuchi}, {and} \bibinfo{person}{Jiro Katto}.}
  \bibinfo{year}{2020}\natexlab{}.
\newblock \showarticletitle{Learned Image Compression with Discretized Gaussian
  Mixture Likelihoods and Attention Modules}. In
  \bibinfo{booktitle}{\emph{CVPR}}.
\newblock


\bibitem[Choi et~al\mbox{.}(2019)]%
        {choi2019variable}
\bibfield{author}{\bibinfo{person}{Yoojin Choi}, \bibinfo{person}{Mostafa
  El-Khamy}, {and} \bibinfo{person}{Jungwon Lee}.}
  \bibinfo{year}{2019}\natexlab{}.
\newblock \showarticletitle{Variable Rate Deep Image Compression with a
  Conditional Autoencoder}. In \bibinfo{booktitle}{\emph{ICCV}}.
\newblock


\bibitem[Company(1999)]%
        {kodak}
\bibfield{author}{\bibinfo{person}{Eastman~Kodak Company}.}
  \bibinfo{year}{1999}\natexlab{}.
\newblock \bibinfo{title}{Kodak Lossless True Color Image Suite}.
\newblock
\newblock
\urldef\tempurl%
\url{http://r0k.us/graphics/kodak/}
\showURL{%
\tempurl}


\bibitem[Cui et~al\mbox{.}(2021)]%
        {cui2021asymmetric}
\bibfield{author}{\bibinfo{person}{Ze Cui}, \bibinfo{person}{Jing Wang},
  \bibinfo{person}{Shangyin Gao}, \bibinfo{person}{Tiansheng Guo},
  \bibinfo{person}{Yihui Feng}, {and} \bibinfo{person}{Bo Bai}.}
  \bibinfo{year}{2021}\natexlab{}.
\newblock \showarticletitle{Asymmetric Gained Deep Image Compression with
  Continuous Rate Adaptation}. In \bibinfo{booktitle}{\emph{CVPR}}.
\newblock


\bibitem[Dinh et~al\mbox{.}(2015)]%
        {dinh2014nice}
\bibfield{author}{\bibinfo{person}{Laurent Dinh}, \bibinfo{person}{David
  Krueger}, {and} \bibinfo{person}{Yoshua Bengio}.}
  \bibinfo{year}{2015}\natexlab{}.
\newblock \showarticletitle{NICE: Non-Linear Independent Components
  Estimation}. In \bibinfo{booktitle}{\emph{ICLRW}}.
\newblock


\bibitem[Dinh et~al\mbox{.}(2017)]%
        {dinh2016density}
\bibfield{author}{\bibinfo{person}{Laurent Dinh}, \bibinfo{person}{Jascha
  Sohl-Dickstein}, {and} \bibinfo{person}{Samy Bengio}.}
  \bibinfo{year}{2017}\natexlab{}.
\newblock \showarticletitle{Density Estimation Using Real NVP}. In
  \bibinfo{booktitle}{\emph{ICLR}}.
\newblock


\bibitem[Duda(2013)]%
        {duda2013asymmetric}
\bibfield{author}{\bibinfo{person}{Jarek Duda}.}
  \bibinfo{year}{2013}\natexlab{}.
\newblock \showarticletitle{Asymmetric Numeral Systems: Entropy Coding
  Combining Speed of Huffman Coding with Compression Rate of Arithmetic
  Coding}.
\newblock \bibinfo{journal}{\emph{arXiv preprint arXiv:1311.2540}}
  (\bibinfo{year}{2013}).
\newblock


\bibitem[Google(2010)]%
        {webp}
\bibfield{author}{\bibinfo{person}{Google}.} \bibinfo{year}{2010}\natexlab{}.
\newblock \bibinfo{title}{Web Picture Format}.
\newblock
\newblock
\urldef\tempurl%
\url{https://chromium.googlesource.com/webm/libwebp}
\showURL{%
\tempurl}


\bibitem[Guo et~al\mbox{.}(2020)]%
        {guo20203}
\bibfield{author}{\bibinfo{person}{Zongyu Guo}, \bibinfo{person}{Yaojun Wu},
  \bibinfo{person}{Runsen Feng}, \bibinfo{person}{Zhizheng Zhang}, {and}
  \bibinfo{person}{Zhibo Chen}.} \bibinfo{year}{2020}\natexlab{}.
\newblock \showarticletitle{3-D Context Entropy Model for Improved Practical
  Image Compression}. In \bibinfo{booktitle}{\emph{CVPRW}}.
\newblock


\bibitem[Helminger et~al\mbox{.}(2020)]%
        {helminger2020lossy}
\bibfield{author}{\bibinfo{person}{Leonhard Helminger},
  \bibinfo{person}{Abdelaziz Djelouah}, \bibinfo{person}{Markus Gross}, {and}
  \bibinfo{person}{Christopher Schroers}.} \bibinfo{year}{2020}\natexlab{}.
\newblock \showarticletitle{Lossy Image Compression with Normalizing Flows}. In
  \bibinfo{booktitle}{\emph{CoRR}}.
\newblock


\bibitem[Ho et~al\mbox{.}(2021)]%
        {ho2021anfic}
\bibfield{author}{\bibinfo{person}{Yung-Han Ho}, \bibinfo{person}{Chih-Chun
  Chan}, \bibinfo{person}{Wen-Hsiao Peng}, \bibinfo{person}{Hsueh-Ming Hang},
  {and} \bibinfo{person}{Marek Doma{\'n}ski}.} \bibinfo{year}{2021}\natexlab{}.
\newblock \showarticletitle{ANFIC: Image Compression Using Augmented
  Normalizing Flows}.
\newblock \bibinfo{journal}{\emph{IEEE Open Journal of Circuits and Systems}}
  \bibinfo{volume}{2} (\bibinfo{year}{2021}), \bibinfo{pages}{613--626}.
\newblock


\bibitem[Hu et~al\mbox{.}(2020)]%
        {hu2020coarse}
\bibfield{author}{\bibinfo{person}{Yueyu Hu}, \bibinfo{person}{Wenhan Yang},
  {and} \bibinfo{person}{Jiaying Liu}.} \bibinfo{year}{2020}\natexlab{}.
\newblock \showarticletitle{Coarse-to-Fine Hyper-Prior Modeling for Learned
  Image Compression}. In \bibinfo{booktitle}{\emph{AAAI}}.
\newblock


\bibitem[Hu et~al\mbox{.}(2021)]%
        {hu2021learning}
\bibfield{author}{\bibinfo{person}{Yueyu Hu}, \bibinfo{person}{Wenhan Yang},
  \bibinfo{person}{Zhan Ma}, {and} \bibinfo{person}{Jiaying Liu}.}
  \bibinfo{year}{2021}\natexlab{}.
\newblock \showarticletitle{Learning End-to-End Lossy Image Compression: A
  Benchmark}.
\newblock \bibinfo{journal}{\emph{IEEE Transactions on Pattern Analysis and
  Machine Intelligence}} (\bibinfo{year}{2021}).
\newblock


\bibitem[Huang et~al\mbox{.}(2017)]%
        {huang2017densely}
\bibfield{author}{\bibinfo{person}{Gao Huang}, \bibinfo{person}{Zhuang Liu},
  \bibinfo{person}{Laurens Van Der~Maaten}, {and} \bibinfo{person}{Kilian~Q
  Weinberger}.} \bibinfo{year}{2017}\natexlab{}.
\newblock \showarticletitle{Densely Connected Convolutional Networks}. In
  \bibinfo{booktitle}{\emph{CVPR}}.
\newblock


\bibitem[Iwai et~al\mbox{.}(2021)]%
        {iwai2021fidelity}
\bibfield{author}{\bibinfo{person}{Shoma Iwai}, \bibinfo{person}{Tomo
  Miyazaki}, \bibinfo{person}{Yoshihiro Sugaya}, {and}
  \bibinfo{person}{Shinichiro Omachi}.} \bibinfo{year}{2021}\natexlab{}.
\newblock \showarticletitle{Fidelity-Controllable Extreme Image Compression
  with Generative Adversarial Networks}. In \bibinfo{booktitle}{\emph{ICPR}}.
\newblock


\bibitem[Johnston et~al\mbox{.}(2018)]%
        {Johnston_2018_CVPR}
\bibfield{author}{\bibinfo{person}{Nick Johnston}, \bibinfo{person}{Damien
  Vincent}, \bibinfo{person}{David Minnen}, \bibinfo{person}{Michele Covell},
  \bibinfo{person}{Saurabh Singh}, \bibinfo{person}{Troy Chinen},
  \bibinfo{person}{Sung~Jin Hwang}, \bibinfo{person}{Joel Shor}, {and}
  \bibinfo{person}{George Toderici}.} \bibinfo{year}{2018}\natexlab{}.
\newblock \showarticletitle{Improved Lossy Image Compression With Priming and
  Spatially Adaptive Bit Rates for Recurrent Networks}. In
  \bibinfo{booktitle}{\emph{CVPR}}.
\newblock


\bibitem[(JVET)(2021)]%
        {vvc}
\bibfield{author}{\bibinfo{person}{Joint Video Experts~Team (JVET)}.}
  \bibinfo{year}{2021}\natexlab{}.
\newblock \bibinfo{title}{VVC Official Test Model VTM}.
\newblock
\newblock
\urldef\tempurl%
\url{https://vcgit.hhi.fraunhofer.de/jvet/VVCSoftware_VTM/-/tree/VTM-12.1}
\showURL{%
\tempurl}
\newblock
\shownote{accessed on April 5, 2021}.


\bibitem[Kingma and Ba(2015)]%
        {Kingma2015Adam}
\bibfield{author}{\bibinfo{person}{Diederik~P. Kingma} {and}
  \bibinfo{person}{Jimmy Ba}.} \bibinfo{year}{2015}\natexlab{}.
\newblock \showarticletitle{Adam: {A} Method for Stochastic Optimization}. In
  \bibinfo{booktitle}{\emph{ICLR}}.
\newblock


\bibitem[Kingma and Dhariwal(2018)]%
        {kingma2018glow}
\bibfield{author}{\bibinfo{person}{Durk~P Kingma} {and}
  \bibinfo{person}{Prafulla Dhariwal}.} \bibinfo{year}{2018}\natexlab{}.
\newblock \showarticletitle{Glow: Generative Flow with Invertible 1x1
  Convolutions}. In \bibinfo{booktitle}{\emph{NeurIPS}}.
\newblock


\bibitem[Lee et~al\mbox{.}(2019)]%
        {lee2018context}
\bibfield{author}{\bibinfo{person}{Jooyoung Lee}, \bibinfo{person}{Seunghyun
  Cho}, {and} \bibinfo{person}{Seung-Kwon Beack}.}
  \bibinfo{year}{2019}\natexlab{}.
\newblock \showarticletitle{Context-Adaptive Entropy Model for End-to-End
  Optimized Image Compression}. In \bibinfo{booktitle}{\emph{ICLR}}.
\newblock


\bibitem[Lin et~al\mbox{.}(2014)]%
        {lin2014microsoft}
\bibfield{author}{\bibinfo{person}{Tsung-Yi Lin}, \bibinfo{person}{Michael
  Maire}, \bibinfo{person}{Serge Belongie}, \bibinfo{person}{James Hays},
  \bibinfo{person}{Pietro Perona}, \bibinfo{person}{Deva Ramanan},
  \bibinfo{person}{Piotr Doll{\'a}r}, {and} \bibinfo{person}{C~Lawrence
  Zitnick}.} \bibinfo{year}{2014}\natexlab{}.
\newblock \showarticletitle{Microsoft COCO: Common Objects in Context}. In
  \bibinfo{booktitle}{\emph{ECCV}}.
\newblock


\bibitem[Liu et~al\mbox{.}(2020)]%
        {liu2020unified}
\bibfield{author}{\bibinfo{person}{Jiaheng Liu}, \bibinfo{person}{Guo Lu},
  \bibinfo{person}{Zhihao Hu}, {and} \bibinfo{person}{Dong Xu}.}
  \bibinfo{year}{2020}\natexlab{}.
\newblock \showarticletitle{A Unified End-to-End Framework for Efficient Deep
  Image Compression}.
\newblock \bibinfo{journal}{\emph{arXiv preprint arXiv:2002.03370}}
  (\bibinfo{year}{2020}).
\newblock


\bibitem[Lugmayr et~al\mbox{.}(2020)]%
        {lugmayr2020srflow}
\bibfield{author}{\bibinfo{person}{Andreas Lugmayr}, \bibinfo{person}{Martin
  Danelljan}, \bibinfo{person}{Luc~Van Gool}, {and} \bibinfo{person}{Radu
  Timofte}.} \bibinfo{year}{2020}\natexlab{}.
\newblock \showarticletitle{SRFlow: Learning the Super-Resolution Space with
  Normalizing Flow}. In \bibinfo{booktitle}{\emph{ECCV}}.
\newblock


\bibitem[Ma et~al\mbox{.}(2020)]%
        {ma2020end}
\bibfield{author}{\bibinfo{person}{Haichuan Ma}, \bibinfo{person}{Dong Liu},
  \bibinfo{person}{Ning Yan}, \bibinfo{person}{Houqiang Li}, {and}
  \bibinfo{person}{Feng Wu}.} \bibinfo{year}{2020}\natexlab{}.
\newblock \showarticletitle{End-to-End Optimized Versatile Image Compression
  with Wavelet-Like Transform}.
\newblock \bibinfo{journal}{\emph{IEEE Transactions on Pattern Analysis and
  Machine Intelligence}}  \bibinfo{volume}{44} (\bibinfo{year}{2020}),
  \bibinfo{pages}{1247--1263}.
\newblock


\bibitem[Ma et~al\mbox{.}(2021)]%
        {ma2021afec}
\bibfield{author}{\bibinfo{person}{Yi Ma}, \bibinfo{person}{Yongqi Zhai},
  \bibinfo{person}{Jiayu Yang}, \bibinfo{person}{Chunhui Yang}, {and}
  \bibinfo{person}{Ronggang Wang}.} \bibinfo{year}{2021}\natexlab{}.
\newblock \showarticletitle{AFEC: Adaptive Feature Extraction Modules for
  Learned Image Compression}. In \bibinfo{booktitle}{\emph{ACMMM}}.
\newblock


\bibitem[Mentzer et~al\mbox{.}(2018)]%
        {mentzer2018conditional}
\bibfield{author}{\bibinfo{person}{Fabian Mentzer}, \bibinfo{person}{Eirikur
  Agustsson}, \bibinfo{person}{Michael Tschannen}, \bibinfo{person}{Radu
  Timofte}, {and} \bibinfo{person}{Luc Van~Gool}.}
  \bibinfo{year}{2018}\natexlab{}.
\newblock \showarticletitle{Conditional Probability Models for Deep Image
  Compression}. In \bibinfo{booktitle}{\emph{CVPR}}.
\newblock


\bibitem[Mentzer et~al\mbox{.}(2020)]%
        {mentzer2020high}
\bibfield{author}{\bibinfo{person}{Fabian Mentzer}, \bibinfo{person}{George~D
  Toderici}, \bibinfo{person}{Michael Tschannen}, {and}
  \bibinfo{person}{Eirikur Agustsson}.} \bibinfo{year}{2020}\natexlab{}.
\newblock \showarticletitle{High-Fidelity Generative Image Compression}. In
  \bibinfo{booktitle}{\emph{NeurIPS}}.
\newblock


\bibitem[Minnen et~al\mbox{.}(2018)]%
        {minnen2018joint}
\bibfield{author}{\bibinfo{person}{David Minnen}, \bibinfo{person}{Johannes
  Ball{\'e}}, {and} \bibinfo{person}{George~D Toderici}.}
  \bibinfo{year}{2018}\natexlab{}.
\newblock \showarticletitle{Joint Autoregressive and Hierarchical Priors for
  Learned Image Compression}. In \bibinfo{booktitle}{\emph{NeurIPS}}.
\newblock


\bibitem[Minnen and Singh(2020)]%
        {minnen2020channel}
\bibfield{author}{\bibinfo{person}{David Minnen} {and} \bibinfo{person}{Saurabh
  Singh}.} \bibinfo{year}{2020}\natexlab{}.
\newblock \showarticletitle{Channel-Wise Autoregressive Entropy Models for
  Learned Image Compression}. In \bibinfo{booktitle}{\emph{ICIP}}.
\newblock


\bibitem[Paszke et~al\mbox{.}(2019)]%
        {paszke2019pytorch}
\bibfield{author}{\bibinfo{person}{Adam Paszke}, \bibinfo{person}{Sam Gross},
  \bibinfo{person}{Francisco Massa}, \bibinfo{person}{Adam Lerer},
  \bibinfo{person}{James Bradbury}, \bibinfo{person}{Gregory Chanan},
  \bibinfo{person}{Trevor Killeen}, \bibinfo{person}{Zeming Lin},
  \bibinfo{person}{Natalia Gimelshein}, \bibinfo{person}{Luca Antiga},
  {et~al\mbox{.}}} \bibinfo{year}{2019}\natexlab{}.
\newblock \showarticletitle{Pytorch: An Imperative Style, High-Performance Deep
  Learning Library}. In \bibinfo{booktitle}{\emph{NeurIPS}}.
\newblock


\bibitem[Qian et~al\mbox{.}(2022)]%
        {qian2022entroformer}
\bibfield{author}{\bibinfo{person}{Yichen Qian}, \bibinfo{person}{Ming Lin},
  \bibinfo{person}{Xiuyu Sun}, \bibinfo{person}{Zhiyu Tan}, {and}
  \bibinfo{person}{Rong Jin}.} \bibinfo{year}{2022}\natexlab{}.
\newblock \showarticletitle{Entroformer: A Transformer-Based Entropy Model for
  Learned Image Compression}. In \bibinfo{booktitle}{\emph{ICLR}}.
\newblock


\bibitem[Rabbani(2002)]%
        {rabbani2002jpeg2000}
\bibfield{author}{\bibinfo{person}{Majid Rabbani}.}
  \bibinfo{year}{2002}\natexlab{}.
\newblock \showarticletitle{JPEG2000: Image Compression Fundamentals, Standards
  and Practice}.
\newblock \bibinfo{journal}{\emph{Journal of Electronic Imaging}}
  \bibinfo{volume}{11}, \bibinfo{number}{2} (\bibinfo{year}{2002}),
  \bibinfo{pages}{286}.
\newblock


\bibitem[Rippel and Bourdev(2017)]%
        {rippel2017real}
\bibfield{author}{\bibinfo{person}{Oren Rippel} {and} \bibinfo{person}{Lubomir
  Bourdev}.} \bibinfo{year}{2017}\natexlab{}.
\newblock \showarticletitle{Real-Time Adaptive Image Compression}. In
  \bibinfo{booktitle}{\emph{ICML}}.
\newblock


\bibitem[Song et~al\mbox{.}(2021)]%
        {song2021variable}
\bibfield{author}{\bibinfo{person}{Myungseo Song}, \bibinfo{person}{Jinyoung
  Choi}, {and} \bibinfo{person}{Bohyung Han}.} \bibinfo{year}{2021}\natexlab{}.
\newblock \showarticletitle{Variable-Rate Deep Image Compression through
  Spatially-Adaptive Feature Transform}. In \bibinfo{booktitle}{\emph{ICCV}}.
\newblock


\bibitem[Sun et~al\mbox{.}(2021)]%
        {sun2021interpolation}
\bibfield{author}{\bibinfo{person}{Zhenhong Sun}, \bibinfo{person}{Zhiyu Tan},
  \bibinfo{person}{Xiuyu Sun}, \bibinfo{person}{Fangyi Zhang},
  \bibinfo{person}{Yichen Qian}, \bibinfo{person}{Dongyang Li}, {and}
  \bibinfo{person}{Hao Li}.} \bibinfo{year}{2021}\natexlab{}.
\newblock \showarticletitle{Interpolation Variable Rate Image Compression}. In
  \bibinfo{booktitle}{\emph{ACMMM}}.
\newblock


\bibitem[Theis et~al\mbox{.}(2017)]%
        {theis2017lossy}
\bibfield{author}{\bibinfo{person}{Lucas Theis}, \bibinfo{person}{Wenzhe Shi},
  \bibinfo{person}{Andrew Cunningham}, {and} \bibinfo{person}{Ferenc
  Husz{\'a}r}.} \bibinfo{year}{2017}\natexlab{}.
\newblock \showarticletitle{Lossy Image Compression with Compressive
  Autoencoders}. In \bibinfo{booktitle}{\emph{ICLR}}.
\newblock


\bibitem[Toderici et~al\mbox{.}(2015)]%
        {toderici2015variable}
\bibfield{author}{\bibinfo{person}{George Toderici}, \bibinfo{person}{Sean~M
  O'Malley}, \bibinfo{person}{Sung~Jin Hwang}, \bibinfo{person}{Damien
  Vincent}, \bibinfo{person}{David Minnen}, \bibinfo{person}{Shumeet Baluja},
  \bibinfo{person}{Michele Covell}, {and} \bibinfo{person}{Rahul Sukthankar}.}
  \bibinfo{year}{2015}\natexlab{}.
\newblock \showarticletitle{Variable Rate Image Compression with Recurrent
  Neural Networks}. In \bibinfo{booktitle}{\emph{ICLR}}.
\newblock


\bibitem[Toderici et~al\mbox{.}(2021)]%
        {CLIC2020}
\bibfield{author}{\bibinfo{person}{George Toderici}, \bibinfo{person}{Wenzhe
  Shi}, \bibinfo{person}{Radu Timofte}, \bibinfo{person}{Johannes~Balle
  Lucas~Theis}, \bibinfo{person}{Eirikur Agustsson}, \bibinfo{person}{Nick
  Johnston}, {and} \bibinfo{person}{Fabian Mentzer}.}
  \bibinfo{year}{2021}\natexlab{}.
\newblock \bibinfo{title}{Workshop and Challenge on Learned Image Compression}.
\newblock
\newblock
\urldef\tempurl%
\url{http://www.compression.cc}
\showURL{%
\tempurl}


\bibitem[Toderici et~al\mbox{.}(2017)]%
        {toderici2017full}
\bibfield{author}{\bibinfo{person}{George Toderici}, \bibinfo{person}{Damien
  Vincent}, \bibinfo{person}{Nick Johnston}, \bibinfo{person}{Sung Jin~Hwang},
  \bibinfo{person}{David Minnen}, \bibinfo{person}{Joel Shor}, {and}
  \bibinfo{person}{Michele Covell}.} \bibinfo{year}{2017}\natexlab{}.
\newblock \showarticletitle{Full Resolution Image Compression with Recurrent
  Neural Networks}. In \bibinfo{booktitle}{\emph{CVPR}}.
\newblock


\bibitem[Wallace(1992)]%
        {wallace1992jpeg}
\bibfield{author}{\bibinfo{person}{Gregory~K Wallace}.}
  \bibinfo{year}{1992}\natexlab{}.
\newblock \showarticletitle{The JPEG Still Picture Compression Standard}.
\newblock \bibinfo{journal}{\emph{IEEE Transactions On consumer Electronics}}
  \bibinfo{volume}{38}, \bibinfo{number}{1} (\bibinfo{year}{1992}),
  \bibinfo{pages}{18--34}.
\newblock


\bibitem[Wang et~al\mbox{.}(2022)]%
        {wang2022neural}
\bibfield{author}{\bibinfo{person}{Dezhao Wang}, \bibinfo{person}{Wenhan Yang},
  \bibinfo{person}{Yueyu Hu}, {and} \bibinfo{person}{Jiaying Liu}.}
  \bibinfo{year}{2022}\natexlab{}.
\newblock \showarticletitle{Neural Data-Dependent Transform for Learned Image
  Compression}.
\newblock \bibinfo{journal}{\emph{arXiv preprint arXiv:2203.04963}}
  (\bibinfo{year}{2022}).
\newblock


\bibitem[Wang et~al\mbox{.}(2018)]%
        {wang2018recovering}
\bibfield{author}{\bibinfo{person}{Xintao Wang}, \bibinfo{person}{Ke Yu},
  \bibinfo{person}{Chao Dong}, {and} \bibinfo{person}{Chen~Change Loy}.}
  \bibinfo{year}{2018}\natexlab{}.
\newblock \showarticletitle{Recovering Realistic Texture in Image
  Super-Resolution by Deep Spatial Feature Transform}. In
  \bibinfo{booktitle}{\emph{CVPR}}.
\newblock


\bibitem[Wu et~al\mbox{.}(2020)]%
        {wu2020gan}
\bibfield{author}{\bibinfo{person}{Lirong Wu}, \bibinfo{person}{Kejie Huang},
  {and} \bibinfo{person}{Haibin Shen}.} \bibinfo{year}{2020}\natexlab{}.
\newblock \showarticletitle{A Gan-Based Tunable Image Compression System}. In
  \bibinfo{booktitle}{\emph{WACV}}.
\newblock


\bibitem[Xiao et~al\mbox{.}(2020)]%
        {xiao2020invertible}
\bibfield{author}{\bibinfo{person}{Mingqing Xiao}, \bibinfo{person}{Shuxin
  Zheng}, \bibinfo{person}{Chang Liu}, \bibinfo{person}{Yaolong Wang},
  \bibinfo{person}{Di He}, \bibinfo{person}{Guolin Ke}, \bibinfo{person}{Jiang
  Bian}, \bibinfo{person}{Zhouchen Lin}, {and} \bibinfo{person}{Tie-Yan Liu}.}
  \bibinfo{year}{2020}\natexlab{}.
\newblock \showarticletitle{Invertible Image Rescaling}. In
  \bibinfo{booktitle}{\emph{ECCV}}.
\newblock


\bibitem[Xie et~al\mbox{.}(2021)]%
        {xie2021enhanced}
\bibfield{author}{\bibinfo{person}{Yueqi Xie}, \bibinfo{person}{Ka~Leong
  Cheng}, {and} \bibinfo{person}{Qifeng Chen}.}
  \bibinfo{year}{2021}\natexlab{}.
\newblock \showarticletitle{Enhanced Invertible Encoding for Learned Image
  Compression}. In \bibinfo{booktitle}{\emph{ACMMM}}.
\newblock


\bibitem[Yang et~al\mbox{.}(2020)]%
        {yang2020variable}
\bibfield{author}{\bibinfo{person}{Fei Yang}, \bibinfo{person}{Luis Herranz},
  \bibinfo{person}{Joost Van De~Weijer}, \bibinfo{person}{Jos{\'e} A~Iglesias
  Guiti{\'a}n}, \bibinfo{person}{Antonio~M L{\'o}pez}, {and}
  \bibinfo{person}{Mikhail~G Mozerov}.} \bibinfo{year}{2020}\natexlab{}.
\newblock \showarticletitle{Variable Rate Deep Image Compression with Modulated
  Autoencoder}.
\newblock \bibinfo{journal}{\emph{IEEE Signal Processing Letters}}
  \bibinfo{volume}{27} (\bibinfo{year}{2020}), \bibinfo{pages}{331--335}.
\newblock


\bibitem[Zhang et~al\mbox{.}(2019)]%
        {zhang2019residual}
\bibfield{author}{\bibinfo{person}{Yulun Zhang}, \bibinfo{person}{Kunpeng Li},
  \bibinfo{person}{Kai Li}, \bibinfo{person}{Bineng Zhong}, {and}
  \bibinfo{person}{Yun Fu}.} \bibinfo{year}{2019}\natexlab{}.
\newblock \showarticletitle{Residual Non-Local Attention Networks for Image
  Restoration}. In \bibinfo{booktitle}{\emph{ICLR}}.
\newblock


\bibitem[Zhou et~al\mbox{.}(2019)]%
        {zhou2019end}
\bibfield{author}{\bibinfo{person}{Lei Zhou}, \bibinfo{person}{Zhenhong Sun},
  \bibinfo{person}{Xiangji Wu}, {and} \bibinfo{person}{Junmin Wu}.}
  \bibinfo{year}{2019}\natexlab{}.
\newblock \showarticletitle{End-to-End Optimized Image Compression with
  Attention Mechanism}. In \bibinfo{booktitle}{\emph{CVPRW}}.
\newblock


\end{thebibliography}
%%
%% The next two lines define the bibliography style to be used, and
%% the bibliography file.

%\clearpage
\newpage

\end{document}